\documentclass[sigconf, authorversion]{acmart}
\AtBeginDocument{%
  \providecommand\BibTeX{{%
    \normalfont B\kern-0.5em{\scshape i\kern-0.25em b}\kern-0.8em\TeX}}}

\usepackage{enumitem}
\usepackage{soul}
\usepackage{float}
\usepackage{xcolor}
\usepackage{subfig}
\setcopyright{acmcopyright}
\copyrightyear{2018}
\acmYear{2018}
\acmDOI{XXXXXXX.XXXXXXX}

\acmConference[Conference acronym 'XX]{Make sure to enter the correct
  conference title from your rights confirmation emai}{June 03--05,
  2018}{Woodstock, NY}
\acmBooktitle{Woodstock '18: ACM Symposium on Neural Gaze Detection,
 June 03--05, 2018, Woodstock, NY} 
\acmPrice{15.00}
\acmISBN{978-1-4503-XXXX-X/18/06}

\begin{document}

\title[Student and educator perspectives in undergraduate engineering universities]{\emph{"With Great Power Comes Great Responsibility!"}: Student and Instructor Perspectives on the influence of LLMs on Undergraduate Engineering Education}

% \affiliation{%

\author{Ishika Joshi}
\email{ishika19310@iiitd.ac.in }
\affiliation{%
  \institution{IIIT Delhi}
  % \streetaddress{Okhla Industrial Estate, Phase 3}
  \city{New Delhi}
  % \state{Delhi}
  \country{India}
  % \postcode{110020}
}

\author{Ritvik Budhiraja}
\email{ritvik19322@iiitd.ac.in}
\affiliation{%
  \institution{IIIT Delhi}
  % \streetaddress{Okhla Industrial Estate, Phase 3}
  \city{New Delhi}
  % \state{Delhi}
  \country{India}
  % \postcode{110020}
}

\author{Pranav Deepak Tanna}
\email{f20212685@pilani.bits-pilani.ac.in}
\affiliation{%
  \institution{BITS Pilani}
  % \streetaddress{Okhla Industrial Estate, Phase 3}
  \city{Pilani}
  % \state{Delhi}
  \country{India}
  % \postcode{110020}
}

\author{Lovenya Jain}
\email{f20211732@pilani.bits-pilani.ac.in}
\affiliation{%
  \institution{BITS Pilani}
  % \streetaddress{Okhla Industrial Estate, Phase 3}
  \city{Pilani}
  % \state{Delhi}
  \country{India}
  % \postcode{110020}
}

\author{Mihika Deshpande}
\email{f20212435@pilani.bits-pilani.ac.in}
\affiliation{%
  \institution{BITS Pilani}
  % \streetaddress{Okhla Industrial Estate, Phase 3}
  \city{Pilani}
  % \state{Delhi}
  \country{India}
  % \postcode{110020}
}

\author{Arjun Srivastava}
\email{f20212490@pilani.bits-pilani.ac.in}
\affiliation{%
  \institution{BITS Pilani}
  % \streetaddress{Okhla Industrial Estate, Phase 3}
  \city{Pilani}
  % \state{Delhi}
  \country{India}
  % \postcode{110020}
}

\author{Srinivas Rallapalli}
\email{r.srinivas@pilani.bits-pilani.ac.in}
\affiliation{%
  \institution{BITS Pilani}
  % \streetaddress{Okhla Industrial Estate, Phase 3}
  \city{Pilani}
  % \state{Delhi}
  \country{India}
  % \postcode{110020}
}

\author{Harshal D Akolekar}
\email{harshal.akolekar@iitj.ac.in}
\affiliation{%
  \institution{IIT Jodhpur}
  % \streetaddress{Okhla Industrial Estate, Phase 3}
  \city{Jodhpur}
  % \state{Delhi}
  \country{India}
  % \postcode{110020}
}

\author{Jagat Sesh Challa}
\email{jagatsesh@pilani.bits-pilani.ac.in}
\affiliation{%
  \institution{BITS Pilani}
  % \streetaddress{Okhla Industrial Estate, Phase 3}
  \city{Pilani}
  % \state{Delhi}
  \country{India}
  % \postcode{110020}
}

\author{Dhruv Kumar}
\email{dhruv.kumar@iiitd.ac.in}
\affiliation{%
  \institution{IIIT Delhi}
  % \streetaddress{Okhla Industrial Estate, Phase 3}
  \city{New Delhi}
  % \state{Delhi}
  \country{India}
  % \postcode{110020}
}

\renewcommand{\shortauthors}{Ishika Joshi, Ritvik Budhiraja et al.}

\begin{abstract}
The rise in popularity of Large Language Models (LLMs) has prompted discussions in academic circles, with students exploring LLM-based tools for coursework inquiries and instructors exploring them for teaching and research. Even though a lot of work is underway to create LLM-based tools tailored for students and instructors, there is a lack of comprehensive user studies that capture the perspectives of students and instructors regarding LLMs. This paper addresses this gap by conducting surveys and interviews within undergraduate engineering universities in India. Using 1306 survey responses among students, 112 student interviews, and 27 instructor interviews around the academic usage of ChatGPT (a popular LLM), this paper offers insights into the current usage patterns, perceived benefits, threats, and challenges, as well as recommendations for enhancing the adoption of LLMs among students and instructors. These insights are further utilized to discuss the practical implications of LLMs in undergraduate engineering education and beyond.
\end{abstract}




\received{20 February 2007}
\received[revised]{12 March 2009}
\received[accepted]{5 June 2009}

\maketitle

\section{Introduction}\label{sec:intro}
% ChatGPT, its popularity, talk about LLMs and their main functions, their relevance to multiple domains including education.
Large Language Models (LLMs) \cite{floridi2020gpt}
like GPT 3.5 \cite{chatgpt} \& GPT-4 
\cite{jagatgptfour} 
and 
Llama 2 \cite{llama2} 
have gained immense popularity in recent times due to their remarkable proficiency in understanding and generating human-like language. These models have been trained on large quantities of data available on the internet with billions of parameters. Their exceptional natural language processing capabilities make them highly valuable for chatbots, virtual assistants, and content creation. These LLMs are poised to have a substantial influence across various domains, including but not limited to healthcare \cite{Reddy2023}, education \cite{Milano2023}, legal services \cite{Nick2023}, software development \cite{MacNeil2023CodeExplain} and finance \cite{Deng2023}.

% LLM in HCI - what kind of research - people have looked at!
Human-Computer Interaction (HCI) scholarship has explored various applications of LLMs. These include use of LLMs for mental healthcare interventions \cite{Jo2023}, content and code generation for research and education \cite{Jiang2022, Liu2023, Petridis2023b, Lee2022, McNutt2023}, prompt design \cite{Pereira2023, Liu2022, Dang2023} and creating interactive applications, and games\cite{Wang2023a, Ashby2023}. Some studies have explored their role in influencing opinions \cite{Jakesch2023}, task management \cite{Arakawa2023}, and long form writing \cite{Mirowski2023} and design of devices for users of augmentative and alternative communication \cite{Valencia2023}.
% Introduce the Problem Statement, why it is important, what all prior work has been already done in this regard.

% In this paper, we analyze the impact of LLMs in undergraduate engineering education in India. We use ChatGPT \cite{chatgpt}, one of the most popular LLM-based Chatbot made publicly available by OpenAI, as our representative LLM-based Chatbot. ChatGPT is built on top of GPT-3.5 and GPT-4.0 LLM models owned by OpenAI. Ever since ChatGPT has been made publicly available, there has been a lot of discussion in the academic community regarding the appropriate and inappropriate use of ChatGPT \hl{[REF]}. There is growing research on how can LLM-based tools like ChatGPT can be integrated into the learning and education process in various contexts within the education domain \hl{[REF]}. Even though there is a lot of research and development ongoing for building LLM-based tools for academic purposes, there are very limited studies on the actual perception of students and instructors regarding ChatGPT or LLMs in general. This paper aims to fill this gap.
Since the public release of ChatGPT, it has sparked extensive discussions within the academic community concerning its appropriate usage \cite{kissinger_opinion_2023, huang_alarmed_2023, chantiri_i_2023, noauthor_people_2023}. There is a growing body of research testing the integration of LLM-based tools like ChatGPT into various educational contexts\cite{Denny2023CopilotCS1, finnie-ansley2023CodexCS2, Reeves2023Parsons, MacNeil2023CodeExplain}. However, there remains a notable dearth of studies addressing the perceptions of students and instructors regarding ChatGPT or LLMs in general within academic environments. We seek to bridge this gap in understanding by investigating these perspectives. This paper conducts an analysis of the influence of Large Language Models (LLMs) in undergraduate engineering education in India. We have selected ChatGPT \cite{chatgpt}, a widely used LLM-based Chatbot developed by OpenAI, as our representative LLM-based Chatbot. ChatGPT is constructed upon the GPT-3.5 \cite{chatgpt} and GPT-4.0 \cite{jagatgptfour} LLM models, proprietary to OpenAI. 

% Talk briefly about methodology
Our research employed a mixed-methods approach \cite{mixed_methods} to comprehensively examine the perceptions of ChatGPT. We aimed to understand its \textit{use cases}, \textit{benefits}, \textit{associated risks}, and gather \textit{suggestions for enhancement}. To achieve this, we conducted user studies involving both students and instructors in three esteemed engineering universities in India. Our data collection methods included online surveys and one-to-one interviews. The data collection process leads to a substantial dataset, comprising \textbf{1306 survey responses from students} and \textbf{112 individual student interviews}. Additionally, we conducted \textbf{27 interviews with instructors}. These interviews and surveys covered a diverse range of engineering disciplines, including but not limited to \textit{computer science}, \textit{electrical engineering}, \textit{electronics engineering}, \textit{mechanical engineering}, \textit{civil engineering}, \textit{chemical engineering} and a few others that include \textit{bioengineering}, \textit{pharmacuetical engineering} and \textit{human-centric design}. This comprehensive approach enabled us to gain insights into ChatGPT's usage across various engineering domains.

% Talk about the research questions
This paper addresses several key research questions to comprehensively investigate the role and impact of ChatGPT in the context of undergraduate engineering education in Indian universities:
\begin{itemize}
    \item \textbf{RQ1:} What are the usage patterns of ChatGPT among students and instructors and their perceived benefits?
    \item \textbf{RQ2:} What are the challenges faced by students and instructors in utilizing ChatGPT in education contexts?
    \item \textbf{RQ3:} How do instructors perceive the influence of ChatGPT on undergraduate education?
    \item \textbf{RQ4:} In what ways can ChatGPT be leveraged to assist instructors and students in enhancing student learning and growth?
\end{itemize}
% \begin{itemize}
%     \item \textbf{RQ1:} To what extent are undergraduate engineering students in Indian universities utilizing ChatGPT, and what are the challenges they encounter? Additionally, what recommendations do students have for enhancing the utility of ChatGPT?
%     \item \textbf{RQ2:} What is the level of awareness among instructors regarding students' utilization of ChatGPT?
%     \item \textbf{RQ3:} How do instructors perceive ChatGPT, both as a beneficial tool for students and as a potential threat to their educational experience?
%     \item \textbf{RQ4:} To what degree is ChatGPT expected to influence student assessments and evaluations within the undergraduate engineering education context?
%     \item \textbf{RQ5:} In what ways can ChatGPT assist instructors in enhancing their teaching methods and effectiveness?
% \end{itemize}

% Talk briefly about the main results
To the best of our knowledge, our research is the first comprehensive user study covering student and instructor perspectives on the impact of LLMs on undergraduate engineering education. Our research findings indicate that students are leveraging ChatGPT for a wide array of purposes, including acquiring rapid information, enhancing their understanding of subjects, summarizing content, and, in some cases, employing it directly to solve coursework assignments. Our investigation also details instructors' apprehensions regarding ChatGPT which is seen to serve as both a valuable tool and a potential academic concern. We draw recommendations from our findings that can be employed in the design and development of educational and academic tools built upon LLMs in the future.

% How the rest of the paper is structured?
The rest of the paper is structured as follows: \S \ref{sec:related_work} provides an overview of existing research at the intersection of Large Language Models (LLMs), Human-Computer Interaction (HCI), and Education. \S \ref{sec:methods} provides the relevant details of our chosen methodology for conducting this study. \S \ref{sec:eval}, presents the evaluation and analysis of the data gathered during the research. \S \ref{sec:discuss} delves into a comprehensive exploration of the various implications arising from the research findings. The paper concludes with \S \ref{sec:conclusion}.
\section{Related Work}\label{sec:related_work}
% For each category of papers, 
% 1. what is the broad problem statement these papers are focusing on?
% 2. How our work is different from this existing work - in terms of user population, methods and results?
% Explain each category of papers in 4-5 statements. 
% the same five questions as above

In this section, we present a literature review on various topics that include (i) LLMs x HCI, (ii) AI in Education, (iii) LLMs in Education, (iv) Student and Instructor Perspectives on LLMs. They are described as follows:

\noindent \subsection{LLMs and HCI.}
\textbf{Prompt Engineering.} 
There is a growing body of research which examines the influence of prompt quality on the responses generated by large language models such as GPT-3. Liu et al. \cite{Liu2022} focus on optimizing text prompts for text-to-image generative models, offering design guidelines for better visual outputs. Zamfirescu-Pereira et al. \cite{Pereira2023} explore whether non-experts can effectively use LLMs with a chatbot design tool, finding that participants struggled with prompt design. Wang et al. \cite{Wang2023a} adapt large language models for mobile interfaces, achieving competitive performance without specialized training. Wu et al. \cite{Wu2022} introduce "Chaining" LLM steps, enhancing task outcomes and collaboration. Dang et al. \cite{Dang2023} show that the users strategically combine diegetic prompts (part of the narrative) and non-diegetic prompts (not a part of the narrative) to guide LLMs in their writing process.
% introduced two types of prompts for Large Language Models (LLMs): diegetic prompts (part of the narrative) and non-diegetic prompts (not a part of the narrative). Their research  
Jiang et al. \cite{Jiang2022} show although LLM-based code synthesis tool offers a promising experience to the study participants, they face challenges in understanding the model's syntax and capabilities. Liu et al. \cite{Liu2023} address the issue of guiding non-expert programmers in generating code by providing natural language prompts to LLMs using "grounded abstraction matching" which translates the code back into a systematic and predictable naturalistic utterance, thus, improving users' understanding and code generation in data analysis.

\noindent\textbf{LLM-based Interactive Applications.} Jo et al. \cite{Jo2023} investigate the use of an LLM-based chatbot "CareCall" for aiding socially isolated individuals in public health interventions. Wang et al. \cite{Wang2023} introduce "PopBlends," a system suggesting conceptual blends for pop culture reference images using traditional knowledge extraction and large language models. Arakawa et al. \cite{Arakawa2023} propose using generative models to boost task engagement and reduce distractions among workers with their system, CatAlyst, which offers context-aware interventions. All of the above-mentioned studies showed that the users of these applications had an overall positive experience. At the same time, these studies identified various challenges that need to be addressed for the widespread adoption of these LLM-based applications. 

Jakesch et al. \cite{Jakesch2023} explore the impact of an LLM-powered writing assistant on users' opinions and writings. The study found that the assistant significantly influenced both the content of participants' writing and their subsequent attitudes, highlighting the need for careful monitoring and engineering of opinions embedded in LLMs. McNutt et al. \cite{McNutt2023} explored code assistants in computational notebooks, identifying challenges such as disambiguation in tasks like data visualization, the need for domain-specific tools, and the importance of polite assistants.

% McNutt et al. \cite{McNutt2023} explored code assistants in computational notebooks, identifying challenges such as disambiguation in tasks like data visualization, the need for domain-specific tools, and the importance of polite assistants. 

% CareCall offers a holistic understanding of individuals, reduces public health workload, and alleviates loneliness. However, challenges remain, requiring careful design and deployment for public health support.

% Users found twice as many suggestions with lower cognitive demand, highlighting the benefits of combining large language models and knowledge bases for creative thinking.

% Customizing the assistant's expressions significantly influenced both content and subsequent attitudes, emphasizing the need for careful management of embedded opinions.

% It effectively helps users resume tasks with reduced cognitive load, offering a novel approach to human-AI collaboration for digital well-being.

\noindent\textbf{LLMs and Datasets.} 
% Lee et al \cite{Lee2022} highlighted the potential of using curated interaction datasets to gain insights into LLMs' generative capabilities, particularly in creative and argumentative writing assistance. The authors introduce CoAuthor, a dataset capturing interactions between writers and GPT-3, enabling the examination of GPT-3's language, ideation, and collaboration capabilities. 
H{\"{a}}m{\"{a}}l{\"{a}}inen et al \cite{Hamalainen2023} explored the potential of LLMs in generating synthetic user research data for Human-Computer Interaction (HCI) research. Their research found that synthetic responses could be quite useful for piloting experiments in HCI but it could jeopardize the reliability of crowdsourced self-report data, if misused.
% They used GPT-3 to generate open-ended questionnaire responses about a particular topic.

\noindent\textbf{LLMs and Creativity.} Mirowski et al \cite{Mirowski2023} introduce Dramatron, a hierarchical language model system designed to generate comprehensive scripts including titles, characters, story elements, and dialogue. Dramatron was able to enhance the coherence of long-form creative writing such as scripts and screenplays. Petridis et al \cite{Petridis2023b} introduce AngleKindling, an LLM-based interactive tool to assist journalists in brainstorming various news angles from documents like press releases. AngleKindling was found to be significantly more helpful and less mentally demanding than previous tools. 
% The study also discussed the potential extension of AngleKindling for application in other knowledge-work domains. 
Lee et al \cite{Lee2022} highlight the potential of using curated interaction datasets to gain insights into LLMs' generative capabilities, particularly in creative and argumentative writing assistance. Jones et al. \cite{Jones2023} explored how artists interacted with algorithmically generated performance instructions and 
% . Artists grappled with the algorithm's disregard for human body limits and developed three modes of interaction: agonistic, perfunctory, and agreeable. The study 
concluded that true collaboration with algorithms is impossible due to their limitations in reciprocity, understanding, and consideration for the human body. Ashby et al. \cite{Ashby2023} introduce a framework for procedural content generation (PCG) \cite{procedural} in role-playing games (RPGs) \cite{roleplay} that prioritizes player-centric generative processes. 
% By combining a knowledge base with an LLM, the method creates quests and dialogue that are contextually grounded and fluent, approaching the quality of hand-crafted content. 
The proposed approach has the potential to enhance player experiences and promote co-creative narratives between humans and AI systems in gaming. Chung et al. \cite{Chung2022} introduced TaleBrush, a generative story ideation tool that enables writers to control and make sense of AI-generated stories by using line sketching interactions with a GPT-based language model. 
% An empirical evaluation demonstrated reliable control of story generation and successful utilization of sketching interactions by users to iteratively create stories according to their creative intentions about character fortunes while taking inspiration from AI-generated stories.

\noindent\textbf{LLMs and Accessibility.} Valencia et al \cite{Valencia2023} explored the potential of LLMs in supporting the users of augmentative and alternative communication (AAC) devices. LLM-generated suggestions were found to potentially save time and reduce user effort during communication. However, users emphasized the importance of aligning the LLM-generated phrases with their communication style and preferences.

\noindent \subsection{AI in Education.}
\textbf{Supporting Students using AI.} Wang et al. \cite{Wang2021} focus on designing natural and prolonged conversations with community-facing conversational agents, focusing on an educational context. The research showed that the student perception of the virtual teaching assistant Jill Watson (JW) changes over time and depends on linguistic elements like verbosity, readability, sentiment, diversity, and adaptability. Winker et al. \cite{Winkler2020} 
% explore the effectiveness of integrating an AI conversational agent (named as Sara) in online video lectures which can provide voice and text-based scaffolding. Using Sara, the authors try to 
address the problem of lack of interaction between instructors and learners in online classes having large audiences. The results showed that Sara, an AI conversational agent was able to significantly improve audience learning in programming tasks, thus, enhancing online learning experiences. Ruan et al. \cite{Ruan2019} introduce a conversational agent for teaching factual knowledge which was able to significantly improve recognition and recall of correct answers, despite being more time-consuming. It shows the potential benefits of educational chatbot systems for non-traditional learning settings. Weitekamp et al. \cite{Weitekamp2020} propose a Simulated Learners technique based method to expedite the creation of Intelligent Tutoring Systems (ITSs) \cite{intelliTutor}. The method could enhance model completeness and reduce authoring time as compared to prior approaches.

\noindent \textbf{Supporting Teachers using AI.} Jensen et al. \cite{Jensen2020} propose an automated approach for providing detailed and actionable automated feedback to teachers to improve their discourse
% The proposed approach involved allowing the teachers to easily record high-quality audio from their classes and 
using speech recognition and machine learning methods. 
% to provide moderately accurate feedback on key dimensions of teacher discourse. 
Aslan et al. \cite{Aslan2019} proposed a real-time, multimodal Student Engagement Analytics Technology to aid teachers in providing just-in-time personalized support to students who risk disengagement. 
% The research utilized a multi-method approach, including a quasi-experimental design and case study, to assess the technology's impact. 
Results indicated that the technology positively contributed to teacher's class practices and student engagement.

% The findings have implications for creating adaptive conversational agents and promoting Mutual Theory of Mind in human-AI interactions in communities.

\noindent \subsection{LLMs in Education.}
% \hl{Jagat sir - This is currently focusing on the application of LLMs in computing education. We probably need to find some papers where LLMs have been applied in education (outside computing education).}
% LLM \cite{floridi2020gpt} tools offer diverse benefits to undergraduate computer science students that include - generating code solutions, educational material, starter code, and debugging assistance. Yet, broader adoption requires addressing ethical, bias, and security concerns. 
LLM \cite{floridi2020gpt} tools offer diverse benefits to undergraduate students. Some of the early studies \cite{becker2023ProsAndCons, Denny2023CopilotCS1, Malinka2023Security, Daun2023Software} examined various challenges and opportunities associated with the utilization of AI code generation tools including OpenAI Codex \cite{OpenAI_Codex}, DeepMind AlphaCode \cite{alphacode}, and Amazon CodeWhisperer \cite{codeWhisperer}.
These studies discussed that these LLM tools can be useful in computer science and related disciplines for a variety of purposes such as (1) generating a variety of code solutions for the students to verify their work and improving their code quality during practice (2) generating high-quality learning material such as new programming exercises \cite{sarsa2022AutoGenerate}, code explanations \cite{Leinonen2023CodeExplanation, sarsa2022AutoGenerate, wermelinger2023Copilot, MacNeil2023CodeExplain} and illustrative examples for the instructors to save their precious time and enhance student learning (3) assisting the instructors and students in providing simple explanations to technical concepts, providing starter code for students to get started, and enhancing programming error messages \cite{Leinonen2023ExplainError} to overcome debugging barriers. At the same time, these LLM tools pose a number of ethical issues such as over-indulgence, plagiarism, carbon footprint of training LLMs, bias related to gender, race, emotion, class, structure of names etc, and security. Similar studies in this domain \cite{Finnie-Ansley2022CS1, wermelinger2023Copilot, Savelka2023MCQAndCode, Reeves2023Parsons,finnie-ansley2023CodexCS2, Savelka2023MCQAndCode, Ouh2023Java, Cipriano2023GPT-3OOP} demonstrate that LLMs can solve a significant portion of programming questions effectively, influenced by task complexity and prompt quality. While they generate accurate solutions, students must still cultivate skills like algorithmic thinking, program comprehension, debugging, and communication.
% : (1) Students may use these tools for solving graded assignments in order to score higher than other students in the same class (2) It is also unclear as to how much of the AI-generated content can be allowed and does not fall under "plagiarism" (3)  Training AI models can consume significant energy and thus raises the question of environmental sustainability. Other challenges posed by these tools include: (1) AI-generated code can reflect stereotypes about gender, race, emotion, class, structure of names etc. (2) AI-generated code may not always be secure and thus, may be vulnerable to security attacks. (3) Students may become over-reliant on these tools for generating solutions instead of properly reading the problem statement and thinking about the solution themselves.

% Numerous research studies have been dedicated to assess the accuracy of LLMs, such as OpenAI Codex \cite{OpenAI_Codex}, GPT-3 \cite{gpt3}, ChatGPT (GPT-3.5 and GPT-4) \cite{chatgpt}, in generating solutions for programming assignments across various computer science courses, including CS1 \cite{Denny2023CopilotCS1, Finnie-Ansley2022CS1, wermelinger2023Copilot, Savelka2023MCQAndCode, Reeves2023Parsons}, CS2 \cite{finnie-ansley2023CodexCS2, Savelka2023MCQAndCode}, object-oriented programming \cite{Ouh2023Java, Cipriano2023GPT-3OOP}, software engineering \cite{Daun2023Software}, and computer security \cite{Malinka2023Security}. 
% The accuracy of LLM-generated solutions varied between 55\% - 70\% in these studies. 

The above-mentioned research studies are primarily focused on assisting students in programming-based learning. Additionally, they do not provide a comprehensive perspective of instructors across different engineering disciplines such as electrical, electronics, mechanical, civil etc. Additionally, the above studies along with some other work \cite{healthcare11060887,app13095783, Ahmad2023, Williamson2023, Alaa2023, Moore2023} don't take into account the actual usage of LLMs by students as well as their perspectives and opinions. Our work aims to fill in these gaps.

\subsection{Student and Instructor Perspective on LLMs.}
Since LLMs have gained prominence only in recent years, a limited number of user case studies have explored the viewpoints of students and instructors \cite{ramazan2023, Shoufan2023, smolansky2023} on LLMs. Ramazan et al. \cite{ramazan2023} conduct a study concentrating on the student perspective of using ChatGPT to solve programming assignments within an object-oriented programming course at a Turkish state university. Smolansky et al. \cite{smolansky2023} used online surveys to analyze both student and instructor perspectives concerning the influence of generative AI on online assessments in higher education universities in the US and Australia.These studies possesses several limitations: Some studies only cover programming assignments within a specific course and also lack an instructor perspective while others focus on essay-type and coding-based evaluations in online setting. Skjuve et al. \cite{Marita2023} conducted a questionnaire study with ChatGPT users to understand their good and bad experiences with ChatGPT but does not focus specifically on the academic context. 

In contrast, our research offers a comprehensive examination of the impact of LLMs on undergraduate engineering education as a whole without confining itself to a particular course or assessment type. 
% Our study considers major engineering disciplines, including computer science, electrical engineering, electronics engineering, mechanical engineering, civil engineering, and chemical engineering. 
Furthermore, in addition to surveys, we incorporate interviews with students and instructors, affording us deeper and more nuanced insights compared to existing research.
\section{Methodology}\label{sec:methods}
\subsection{Research Design}\label{sec:research_design} 

We adopted a mixed-methods \cite{mixed_methods} research approach for this study with an exploratory design \cite{noauthor_exploratory_2023}. Using a mixed-methods approach enabled us to leverage qualitative and quantitative methodologies, resulting in a thorough and nuanced understanding of our research problem.
We selected three universities (referred to as University A, University B, and University C in visualizations) in India for our data collection. All three universities focus primarily on higher education and research in engineering and sciences. We focused our user studies on ChatGPT \cite{chatgpt} as it was the most widely used LLM during the time this study was conducted \cite{hu_chatgpt_2023}. The survey observed 1306 responses. We carried out interviews with 112 undergraduate engineering students spread across different academic years at the selected universities. 
Interviews with 27 undergraduate engineering instructors\footnote{By instructor, we refer to assistant professors, associate professors, and professors who teach engineering courses as well as conduct scientific research.} were also carried out in the same universities. 
The students and instructors were recruited through a combination of purposive and convenience sampling \cite{convPurpSampling}. 
The interviews were conducted in either online mode or in person and the audio recordings of these interviews were further analysed. The interviewees provided both written and verbal consent for the interview and the recordings. A small fraction of students and instructors did not give their consent for recordings. In such cases, our research team made notes during the interview which were later utilized for analysis.  All the research materials and protocols for this study were reviewed and approved by our university's Institutional Review Board (IRB).

\begin{figure}[t]
  \centering
  \subfloat[]{\includegraphics[width=0.31\textwidth]{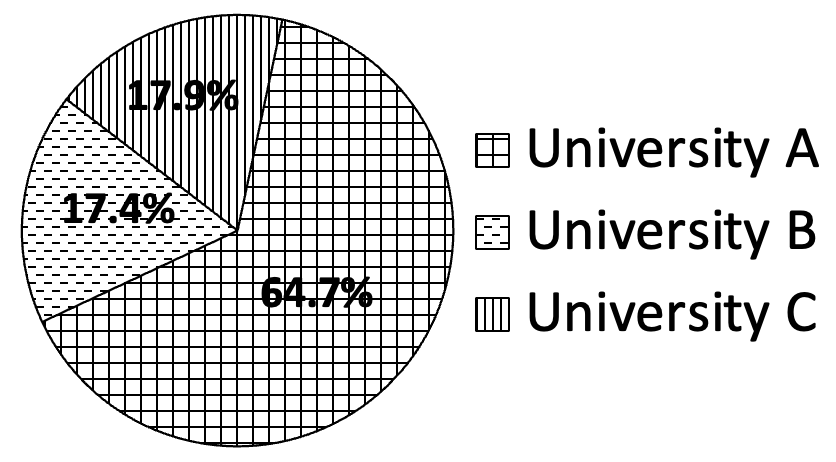}}
  \hspace{0.05cm}
  \subfloat[]{\includegraphics[width=0.45\textwidth]{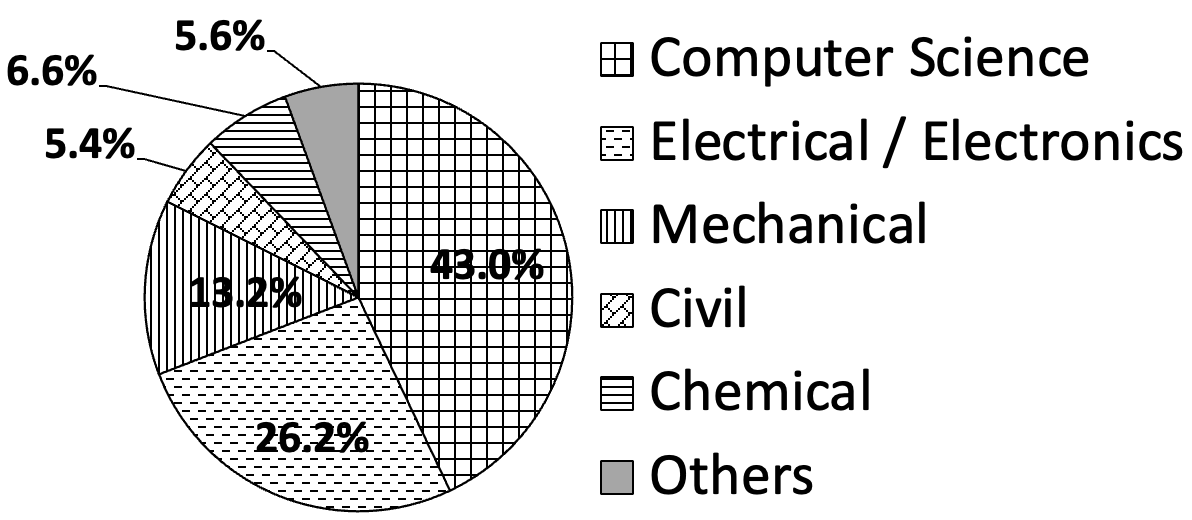}} 
  \hspace{0.05cm}
  % \subfloat[]{\includegraphics[width=0.24\textwidth]{files/new_figs/fig1c.png}}
  \caption{Distribution of the 1306 student survey participants in terms of (a) university, (b) engineering stream}
\label{fig:survey_demograph}
\end{figure}

\subsection{Survey Design}
A survey was made using Google Forms and was circulated through mailing lists of all three universities through authorized personnel. The email explained the purpose of the study and requested students to respond to the form. The survey covered various facets of LLMs, keeping ChatGPT as the focus, within an academic context through 10 questions- including its \textit{frequency of use} (as addressed in two questions), its \textit{popular use cases} (covered by two questions), and \textit{participants' overarching perspectives on the benefits \& drawbacks}, and thoughts regarding \textit{ChatGPT's utility as a tool in engineering education} (encompassing six questions). We addressed these themes to help us establish a foundation of typical student perceptions, attitudes, difficulties, and anticipations related to ChatGPT. The survey consisted of single-choice (5 questions), Likert scales (1 question), and multiple-choice options (3 questions) in addition to two text-based response fields, enabling us to gather both qualitative and quantitative insights. It took around 3-5 minutes to fill the form. Respondents were also asked to mention their university name and their academic year. The respondents were also explained their contribution to the study in the survey introduction along with an assurance of the anonymity of their responses as their names or any such identifiers were not collected. The survey responses were analyzed and subsequently used in the study to obtain insights and shape the design of interview questions. Figure \ref{fig:survey_demograph} shows the distribution of survey participants in terms of university and engineering streams. 

% \begin{figure}
% \centering
% \includegraphics[width=1\textwidth]{demograph.jpg}
% \caption{Distribution of the 1306 UG survey participants in terms of (a) university, (b) year of joining the respective university (c) engineering stream.}
% \label{fig:survey_demograph}
% \end{figure}

% \begin{figure}[t]
%   \centering
%   \subfloat[]{\includegraphics[width=0.24\textwidth]{files/new_figs/fig1a_old.png}}
%   \subfloat[]{\includegraphics[width=0.35\textwidth]{files/new_figs/fig1b_old.png}}  
%   \subfloat[]{\includegraphics[width=0.2\textwidth]{files/new_figs/fig1c_old.png}}
%   \caption{Distribution of the student interview participants in terms of (a) university, (b) year of joining the respective university (c) engineering stream.}
% \label{fig:survey_demograph1}
% \end{figure}

\subsection{Interviews}

The interviews were conducted with two sets of stakeholders - \textit{students} and \textit{instructors} from all three universities, whose details are explained as follows:

\begin{figure}[b]
  \centering
  \subfloat[]{\includegraphics[width=0.32\textwidth]{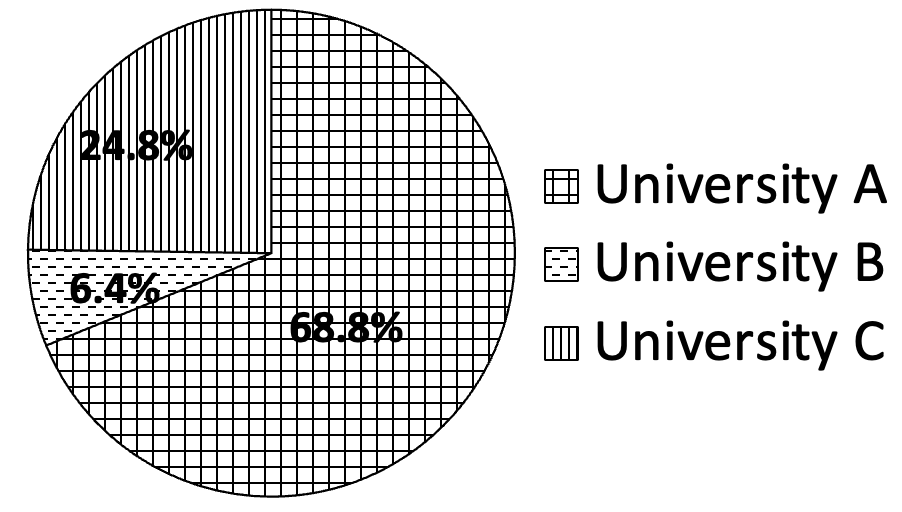}}
  \hspace{0.1cm}
  \subfloat[]{\includegraphics[width=0.39\textwidth]{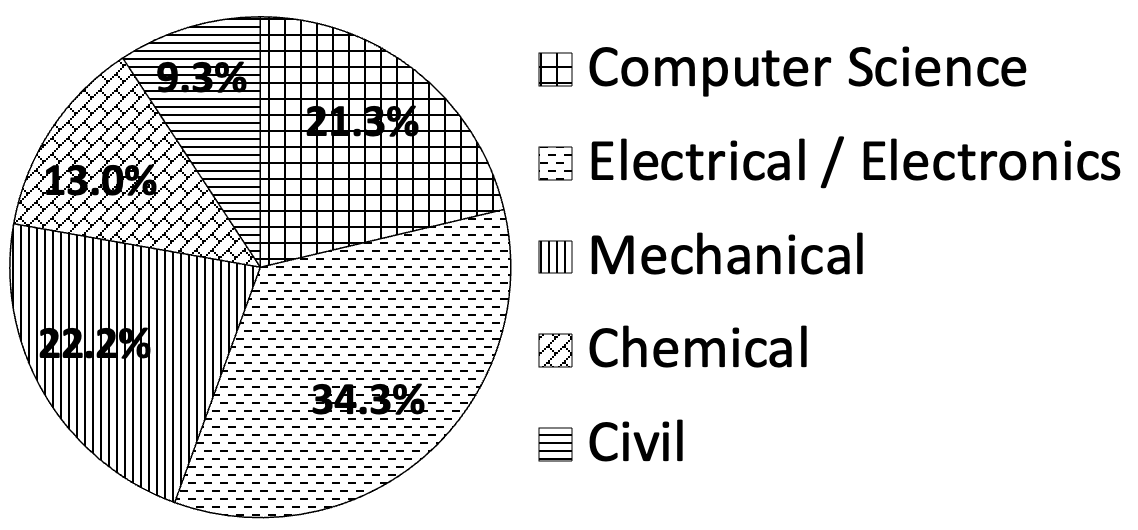}} 
  \hspace{0.05cm}
  % \subfloat[]{\includegraphics[width=0.25\textwidth]{files/new_figs/fig2c.png}}
  \caption{Distribution of 112 student interviews in terms of (a) university (b) engineering stream}
\label{fig:instructor_demograph}
\end{figure}

\subsubsection{\textbf{Student Interviews}}
112 interviews were conducted to comprehensively understand students' experiences using ChatGPT in their academic workflows. This involved exploring their motivations, routines, advantages, obstacles, attitudes, perceptions, and biases related to Language Model Models (LLMs), focusing on ChatGPT. Our sample included students across all four academic years of undergraduate engineering in Indian Universities based on their availability and access to them. The selection criteria for recruiting participants involved being enrolled in an undergraduate engineering program and having used ChatGPT for academic workflows to ensure that the insights gathered come from true experiences. Each interview lasted for an average length of around 10 minutes. Online interviews were conducted and recorded over Google Meet by the research team. In-person interviews were also recorded using Google Meet. Figure 2 shows the distribution of students in terms of university and the engineering stream. There were 86 male participants and 26 female participants in the student interviews which owed to the poor representation of women in the STEM field in India.
% Also, here is the batch-wise split (based on year of admission) of the student participants in the interviews: 3 students from 2019 batch, 9 from 2020 batch, 82 from 2021 batch, and 18 from 2022 batch. 
% \textcolor{red}{WRITE ABOUT THE STREAMS AND NUMBER OF STUDENTS OF EACH STREAM}

% \begin{figure}
% \centering
% \\includegraphics[width=0.5\textwidth]{files/new_figs/temp.png}
% \caption{Distribution of the student interview participants in terms of (a) university, (b) year of joining the respective university (c) engineering stream.}
% \label{fig:survey_demograph}
% \end{figure}

\textit{\textbf{Interview Design.}}
The interviews were semi-structured, designed based on the established research questions, and further refined by incorporating popular insights derived from the initial survey results. The interview consisted of six themes spread over six primary questions- \textit{awareness and familiarity} with ChatGPT, usage patterns, \textit{use cases in academic contexts}, \textit{problems experienced} in using ChatGPT, \textit{potential harm to learning}, and \textit{perceptions and attitudes} towards incorporation of AI in educational settings. Participants were prompted to share how ChatGPT aided their learning and workflows, highlight challenges faced and strategies used, and talk about their fundamental beliefs and opinions, including whether they saw ChatGPT as a harmful or a helpful learning tool.

\subsection{Instructor Interviews}
Interviews were conducted with 27 instructors. The instructors revolved around understanding the experiences of instructors using ChatGPT, their \textit{use cases}, \textit{usage patterns}, \textit{difficulties faced}, \textit{perceptions}, and \textit{attitudes} about LLMs in an academic context. The average length of the interview was 20 minutes. The instructors were selected while ensuring they are regular faculty for undergraduate engineering courses in Indian Universities doing both teaching and research. 9 of the interviews were conducted in person.
Figure \ref{fig:instructor_demograph} shows the distribution of instructors in terms of university and engineering streams. There were 22 male instructors and 5 female instructors. Also, 3 instructors were Professors, 5 were Associate Professors, and 19 were Assistant Professors. This was reflective of the representation of these roles in these colleges. The female representation in STEM education is very low which made it difficult to access more female instructors.

\begin{figure}[b]
  \centering
  \subfloat[]{\includegraphics[width=0.30\textwidth]{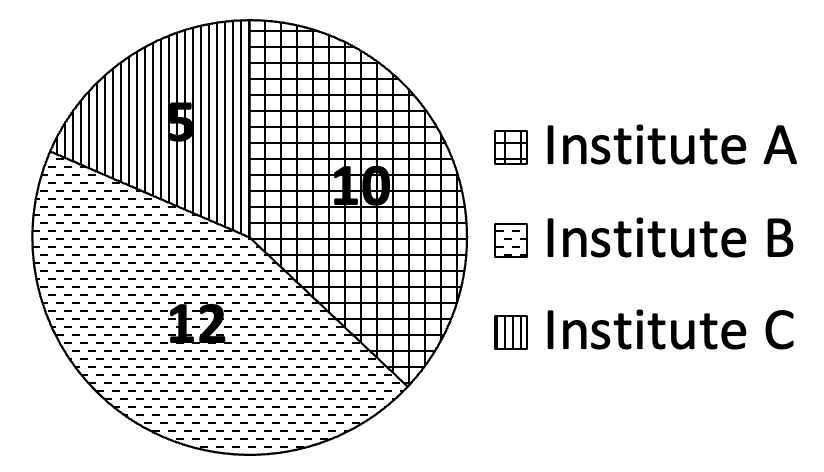}}
  \hspace{0.1cm}
  \subfloat[]{\includegraphics[width=0.39\textwidth]{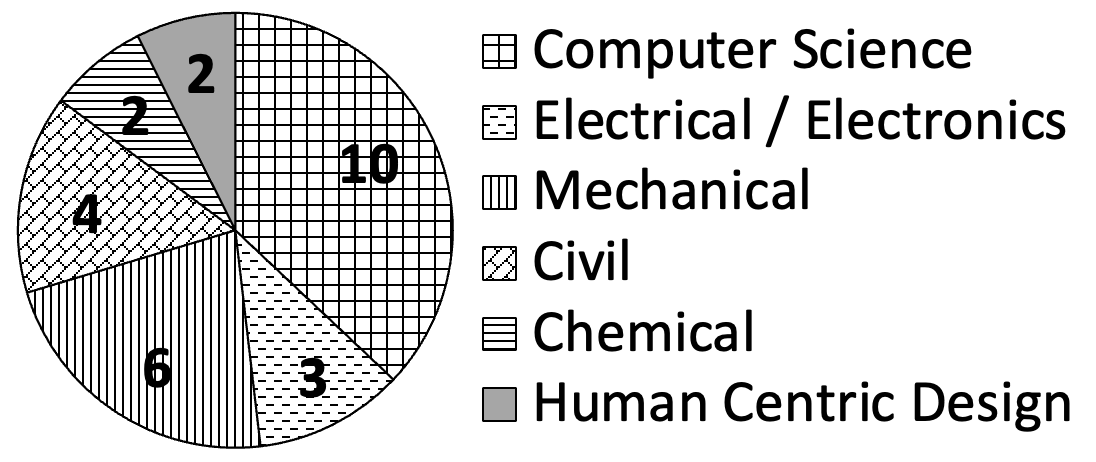}} 
  % \hspace{0.05cm}
  % \subfloat[]{\includegraphics[width=0.28\textwidth]{files/new_figs/fig3a.png}}
  \caption{Distribution of the 27 instructor interviews in terms of (a) university (institute) (b) engineering stream}
\label{fig:instructor_demograph}
\end{figure}

\textit{\textbf{Interview Design.}}
The interviews were semi-structured, and designed based on the established research questions. The interview consisted of six themes spread over six primary questions- awareness and familiarity with ChatGPT, usage patterns, personal use cases in academic contexts along with use cases for their curriculum design, personal problems experienced in using ChatGPT, anticipated harms to student learning, and perceptions and attitudes towards incorporation of AI in educational settings. Participants were prompted to share insights on how ChatGPT impacted their workflows, discuss its pros and cons for student learning and academics, explore effective integration in academic contexts, and talk about their fundamental beliefs and opinions.

\subsection{Data Analysis}
139 interviews (27 instructor interviews, 112 student interviews) were transcribed by the research team. All transcriptions in the Hindi language were translated into English to maintain the language consistency across transcripts. This was followed by a Thematic Analysis approach \cite{braun_using_2006}. Multiple rounds of coding of the transcripts were conducted to extract emerging themes. Following this, the codes were collected on a collaborative platform, FigJam\footnote{https://www.figma.com/figjam/}. The first two authors carried out a three-level analysis of the codes for student interviews by categorizing them into different buckets until the point saturation was achieved. The final categorizations, as explained in the findings section, were - (1) Existing Usage Patterns and Benefits of ChatGPT in academia, (2) Challenges pertaining to ChatGPT usage, and (3) Opportunities and Recommendations for improvement. A similar process was carried out for the instructor interviews, with the insights being segregated into - (1) Instructor Awareness about ChatGPT and its uses, (2) Instructor Perceptions of Student Learning through ChatGPT, (3) Influences on Teaching Methodologies, and (4) Instructor Recommendations.

\subsection{Ethical Considerations}
In conducting this study, we diligently addressed various ethical considerations to maintain transparency and protect the privacy and well-being of our participants. All research materials and procedures were subjected to a detailed review and approval process by our university's Institutional Review Board (IRB). In the surveys, the participants were ensured about the voluntary nature of the survey, the anonymity of their responses, and the purpose of the study. Before engaging in the interviews, every participant was asked to express their consent through a consent form, which explicitly outlined the study's objectives, the voluntary nature of their involvement, and the assurance of anonymity and confidentiality. 
Additionally, we obtained explicit written and verbal consent from participants. All the data collected was anonymized and stored on Google Drive with limited access to some members of the research team. Our research team comprises members with expertise in HCI, interaction design, AI, and engineering education with all authors physically located in India.

\subsection{Limitations}
The study has been structured to exclude any elements specific to individual universities. However, due to resource and time constraints, it was only feasible to include students from a limited number of universities. This limitation has the potential to introduce undisclosed biases related to the academic capabilities and approaches of the universities included, despite having access to a diverse and extensive participant pool.
\section{Evaluation}\label{sec:eval}
\subsection{Student Perspective on ChatGPT - Quantitative Evaluation}
\begin{figure}[h!]
  \centering
  \subfloat[]{\includegraphics[width=0.33\textwidth]{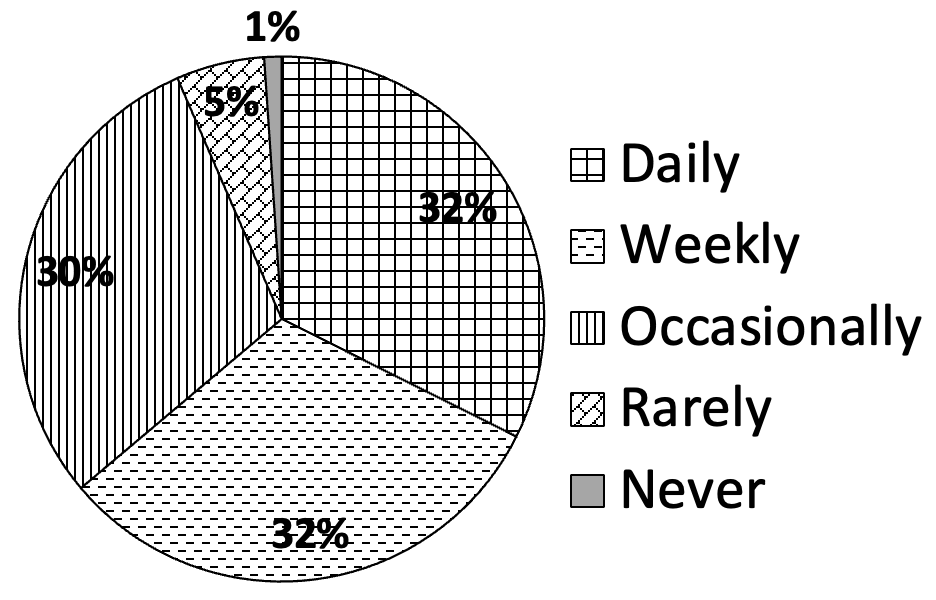}}
  \hspace{0.1cm}
  \subfloat[]{\includegraphics[width=0.33\textwidth]{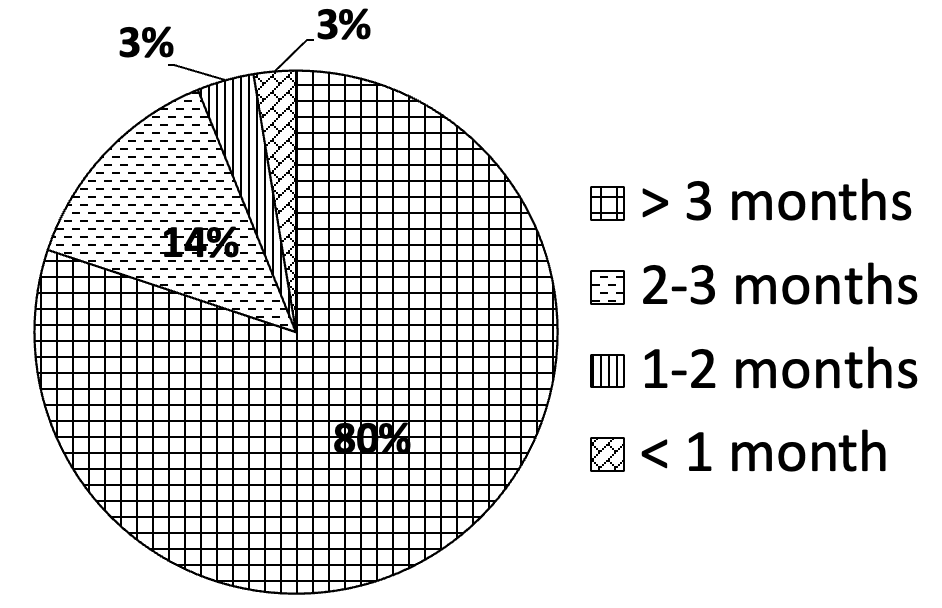}} 
  \hspace{0.1cm}
  \subfloat[]{\includegraphics[width=0.33\textwidth]{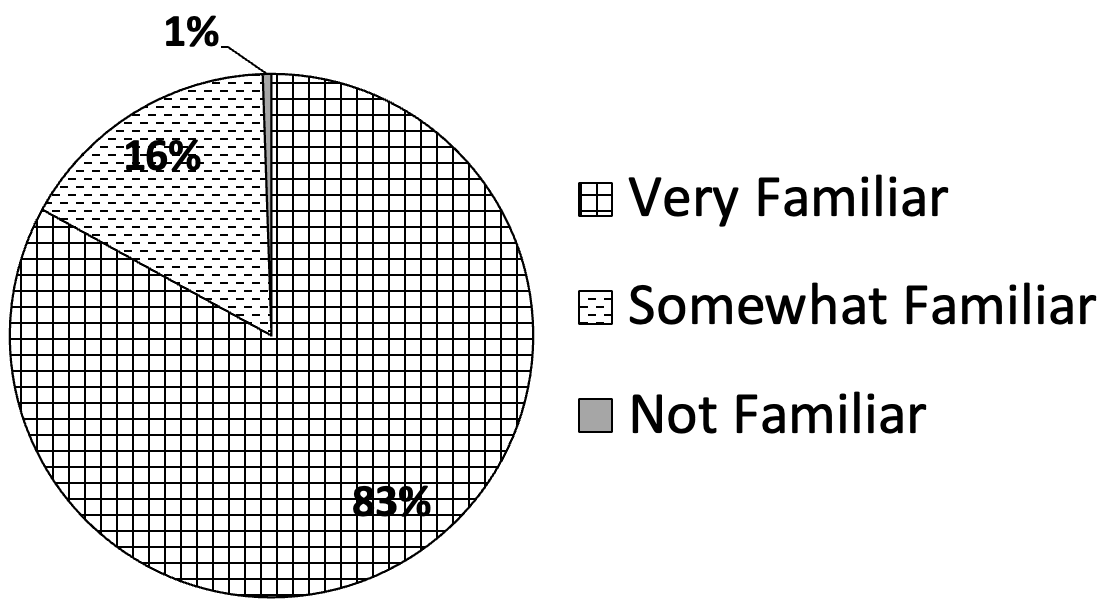}}
  \caption{Illustrating (a) how often do students use ChatGPT? (b) for how long students have been using ChatGPT? (c) how familiar are students with ChatGPT?}
\label{fig:new}
\end{figure}

\begin{figure}[h!]
  \centering
  % % \vspace{0.5cm}
  \subfloat[]{\includegraphics[width=0.45\textwidth]{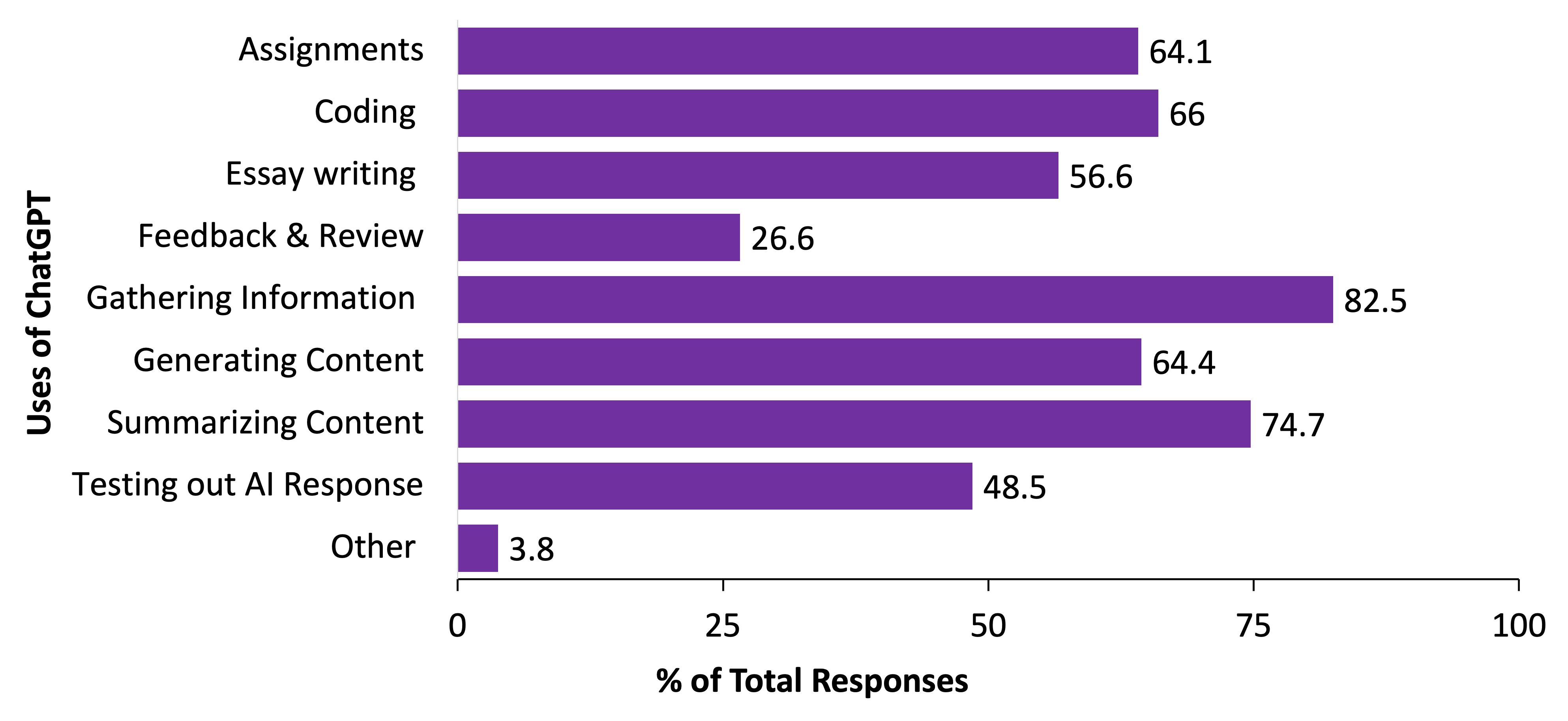}} \\
  % \hspace{0.1cm}
  \subfloat[]{\includegraphics[width=0.45\textwidth]{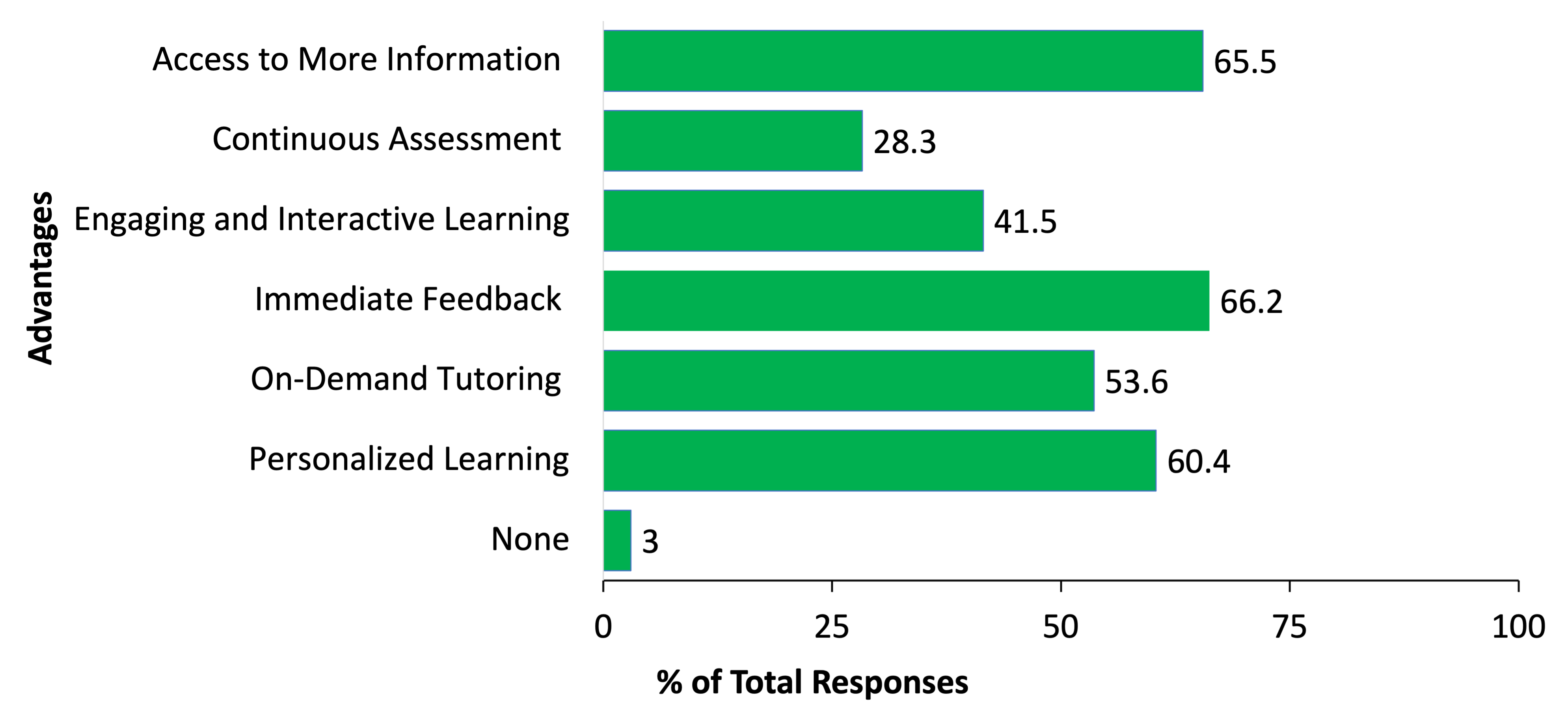}} \\
  % \hspace{0.05cm}
  \subfloat[]{\includegraphics[width=0.45\textwidth]{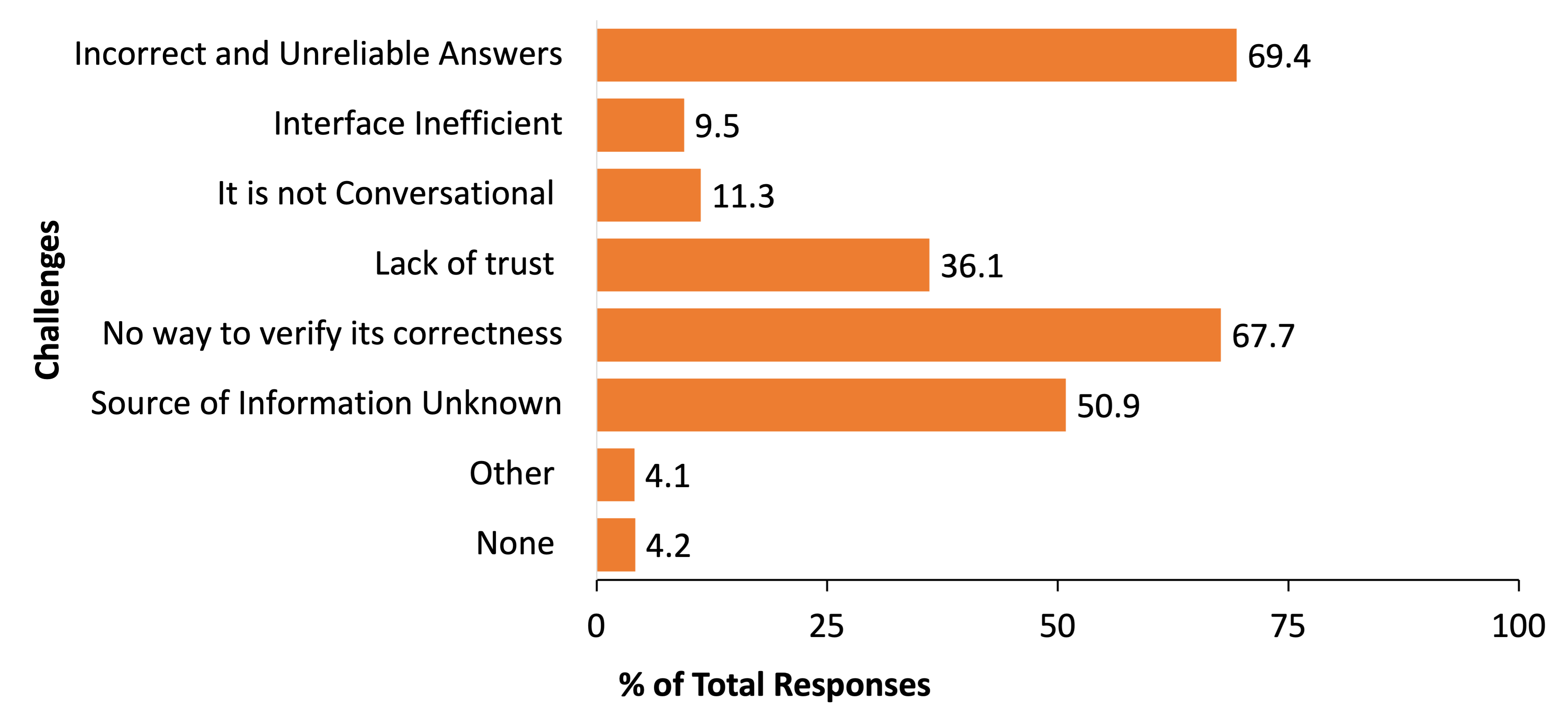}}
  \caption{Perception of ChatGPT based on 1306 student survey responses regarding its (a) usage, (b) advantages, and (c) challenges}
\label{fig:perception}
% \vspace{0.5cm}
\end{figure}

The survey attracted 1306 responses across various engineering disciplines from Undergraduate students across multiple universities as underscored in the methodology section. We received responses from all engineering streams across the chosen universities including, Computer Science Engineering and its allied disciplines, Electrical / Electronics Engineering, Mechanical Engineering, Civil Engineering, Chemical Engineering, and others (Bioengineering, Pharmaceutical Engineering, Human-Centric Design, etc.). In India, an Engineering program is usually 4 years long. In our responses, we got \textbf{33.9\%} responses from the freshmen year, \textbf{27.5\%} from the sophomore year, \textbf{29.3\%} from the junior year, and \textbf{9.3\%} from the senior year. Since the study was conducted during a period of transition to College Graduates for senior year students, there was less engagement reported from them, resulting in fewer responses.

% \begin{figure}
% \centering
% \fbox{\includegraphics[width=1\textwidth]{res1.jpg}}
% \caption{Distribution of the 1306 UG survey participants in terms of (a) the institute, (b) year of joining the respective institution (c) degree.}
% \label{fig:res1}
% \end{figure}

\begin{figure}[h!]
  \centering
  \subfloat[]{\includegraphics[width=0.33\textwidth]{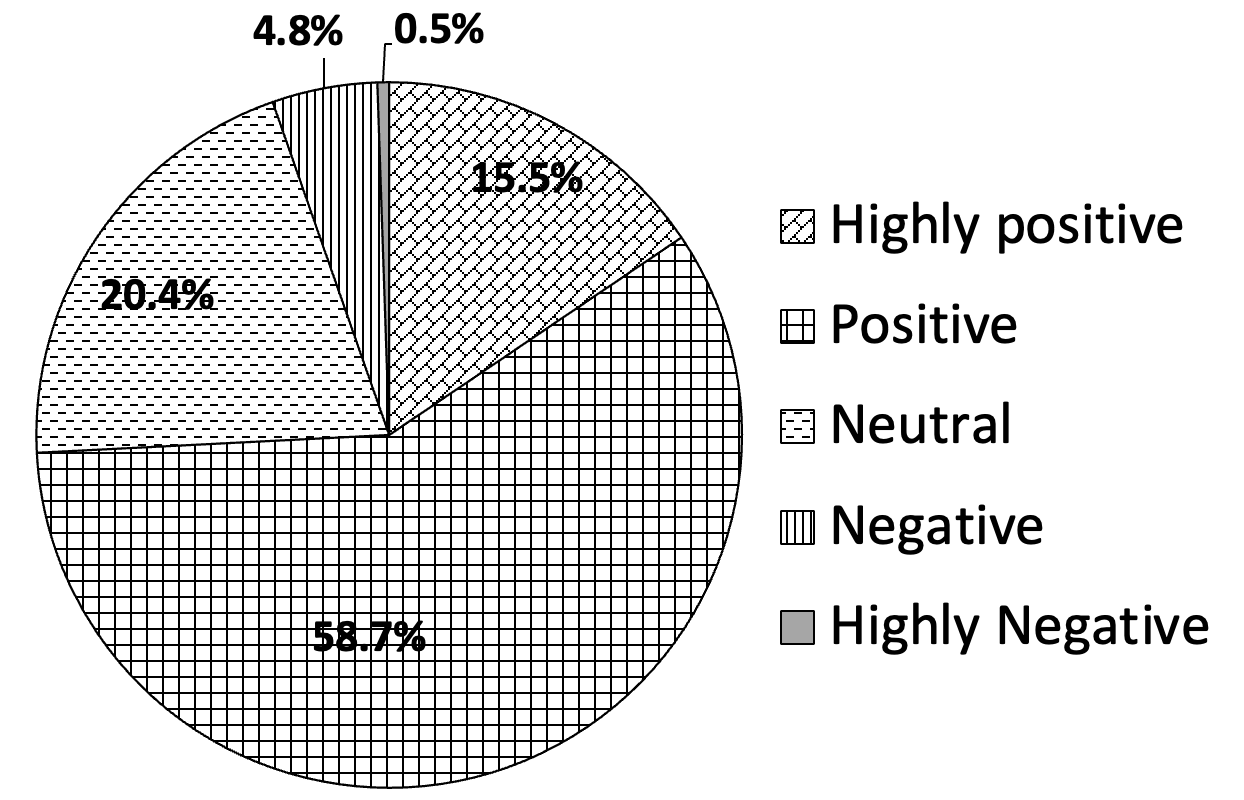}}
  \hspace{0.1cm}
  \subfloat[]{\includegraphics[width=0.33\textwidth]{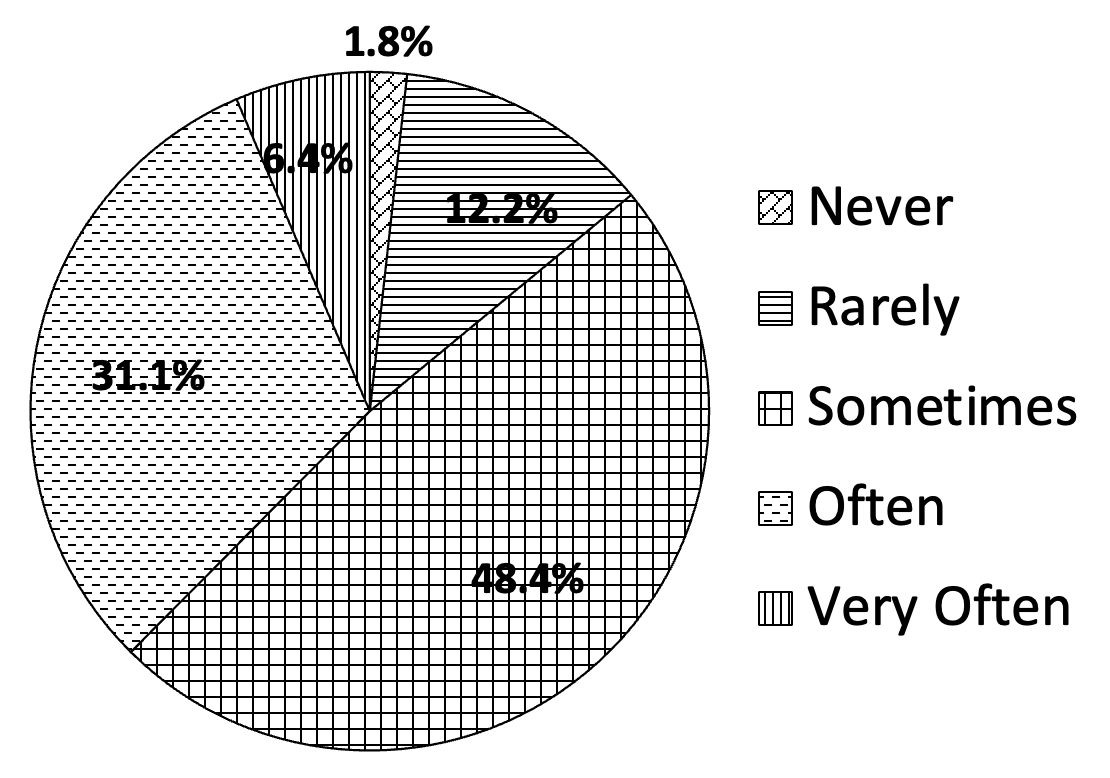}} 
  \hspace{0.05cm}
  \subfloat[]{\includegraphics[width=0.25\textwidth]{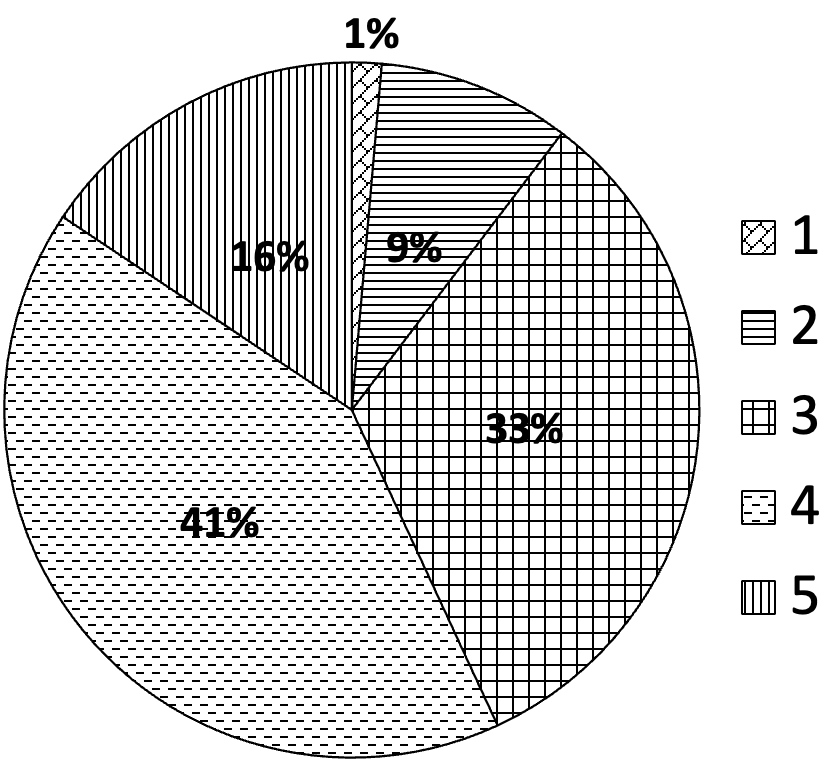}}
  \caption{(a) Overall perception of students towards ChatGPT as a learning and education tool (b) How often do students run into problems using ChatGPT? (c) To what extent do students believe ChatGPT can aid in their coursework-related queries?}
\label{fig:feedback}
\end{figure}

\noindent The results of our study are depicted in Figure \ref{fig:new}. In the study responses, 83\% reported to be \textit{very familiar} with ChatGPT. Meanwhile, 16\% students were \textit{somewhat familiar} and 1\% students reported to be \textit{not familiar}. This reflects its high usage and adoption among undergraduate engineering students. 32\% users claimed to be daily users of ChatGPT, 32\% students use it weekly, 30\% use it occasionally and 5\% of the survey population rarely used ChatGPT. A small percentage (1\%) claimed to have never used it. Additionally, the survey's findings indicated that most of the respondents (80\%) had been using ChatGPT for a significantly long period (greater than 3 months), highlighting the vast retention rate of the service among undergraduate engineering students.
\\
When asked about the most common use cases of AI, the maximum number of respondents (82.5\%) found it useful for gathering information as indicated by Figure \ref{fig:perception}a. Some other highly ranked use cases were summarizing content (74.7\%), assistance in coding (66\%), assistance in assignments (64\%), assistance in generating content (64.3\%), assistance in essay writing (56.5\%) and just testing out the capabilities of the AI (48.5\%). When asked about the extent of usefulness of ChatGPT in regular coursework, 41.2\% students chose level 4 on a Likert scale from 1 - 5, level 1 being not useful at all and level 5 being highly useful. This is indicated by Figure \ref{fig:perception}b. When asked about the potential benefits of integrating LLMs like ChatGPT in academic workflows, 66.2\% respondents found it to be beneficial for immediate feedback, as indicated in Figure \ref{fig:feedback}c. 65.5\% respondents found it to be a source to access more information, 60.4\% found it to enable personalized learning, 53.6\% thought of it as an on-demand tutor, and 41.5\% found it to be an engaging and interactive learning platform. This has been indicated in Figure \ref{fig:perception}b.
\\

\noindent However, almost half the participants (48.4\%) reported to have sometimes faced problems while using ChatGPT, and 31.1\% reported to have often experienced problems using ChatGPT, as indicated in Figure \ref{fig:feedback}b. Maximum number of participants (69.4\%) ranked incorrect and unreliable results as the most common problem. 67.6\% of participants found the absence of a way to verify the information provided by ChatGPT as a common problem, as indicated in Figure \ref{fig:perception}c. While 36.1\% of participants expressed a lack of trust towards the AI. 11.2\% also found it to not be conversational enough. It's worth noting that 4.2\% of the participants reported not having any problems with the AI.
    
\noindent In all, 58.7\% described their overall perception of ChatGPT as a learning and education tool to be positive with 15.5\% reporting a highly positive perception, as indicated by Figure \ref{fig:feedback}a. 20.4\% respondents had a neutral perception while 4.8\% had a negative perception. 0.5\% respondents also had a highly negative perception of ChatGPT as a learning tool.

\subsection{Student Perspective on ChatGPT - Qualitative Evaluation}
\subsubsection{\textbf{Transitioning Workflows - The Advantages of Leveraging ChatGPT}}
Students have used ChatGPT for their academic coursework and other use cases in relation to their academic requirements. Our interviews found that a large percentage of participants seek general assistance from ChatGPT, such as deriving lists of important topics for a course and seeking explanations for various topics, with the advantage of generating as many examples as required. Students have also used ChatGPT for problem solving, such as coding-related help, debugging sections of code, learning new programming languages, creating edge cases and tests for coding-related problems, and learning topics such as Data Structures and Algorithms. Other problem-solving topics include generating detailed literature reviews, paraphrasing, solving numerical-based questions and generating practice questions. While the majority of the students use ChatGPT to deal with problems that they are facing while coming up with solutions to problems, a fair share of students have also admitted to using ChatGPT directly to solve their assignments for them. 
\begin{quote}
    "I'll say it has helped me being efficient because generally when we read books and all, [...] we have to read chapters that are 50 - 60 pages. So to shorten them down to, maybe 10 pages and that helps me study efficiently." -[P63]
\end{quote}

% \begin{figure}
% \centering
% \fbox{\includegraphics[width=0.9\textwidth]{res2.png}}
% \fbox{\includegraphics[width=0.9\textwidth]{res3.jpg}}
% \fbox{\includegraphics[width=0.9\textwidth]{res4.jpg}}
% \caption{xxx }
% \label{fig:res2}
% \end{figure}

Most students favored ChatGPT for quick information retrieval, knowledge enhancement, and summarizing data. Many utilized its content-generative capabilities for information seeking. Some preferred extracting keywords from research papers with ChatGPT rather than reading the entire document, then requesting brief explanations of the extracted keywords. The majority of the students reported that getting summarised content through ChatGPT is a way to save their time and effort. Students found that using ChatGPT streamlined their tasks, eliminating the need to interrupt their workflow to use external resources or conduct online searches. Many students suggested that ChatGPT is capable of writing high-quality theoretical and verbose content, and hence is useful for assignments that require polished, effective use of language. Generating descriptive essays, creative writing tasks, and reports are some popular language-related uses of ChatGPT among students. Additionally, students frequently rely on ChatGPT to obtain concise information for assistive content like slide decks and presentations. Beyond assignments and information, ChatGPT is also utilized for crafting emails and resumes, and obtaining structural content guidelines. Participants have also used ChatGPT to manage schedules, create timetables, and manage coursework in their day-to-day lives. Some also use it to manage internship task management and design project roadmaps or milestones.
\begin{quote}
    "I use it to collect certain information like, quick information, if I don't Google it. But if I [...] want to Know details about a very intricate question, and I want to know the details of it, then I'll just ask ChatGPT." -[P23]
\end{quote}

Doubt solving was a major use case for students, as it avoids the dependence on professors, senior students, and fellow batch mates for catering to certain doubts. In addition to gaining academic independence when it comes to asking questions and clarifying concepts, students also mentioned that they use ChatGPT in parallel to their existing classes, and in some cases use ChatGPT to learn alongside an ongoing class. The majority of the students reported that the conversational nature and interactivity of ChatGPT allow for an easier onboarding experience and make it a good learning tool. Following this, a large percentage of students reported using ChatGPT as a tutor, and utilizing the tool for tasks such as asking questions that are not a part of any existing internet resource, getting a step-by-step explanation for concepts and solutions, and generating practice questions outside the typical textbook question banks. Students mentioned that they are using ChatGPT as an assistant in self-study, and the tool is a very useful starting point for further manual search, aiding their initial ideation. Some of their professors even endorsed its use in class. A fraction of the students also mentioned that outside their usual coursework, they also used ChatGPT to enhance their communication skills such as improving their vocabulary and fluency in English.  
\begin{quote}
    "It has helped me understand new things. Sometimes when I'm stuck at some question or anything it helps to think of it from a new angle or newest perspective ChatGPT comes to rescue." -[P30]
\end{quote}

% Overall, students found that using ChatGPT streamlined their tasks, eliminating the need to interrupt their workflow to use external resources or conduct online searches. Some of their professors even endorsed its use in class. Despite the tool's challenges, many students believed ChatGPT was valuable, saving time and energy and reducing mental blocks in their work.

% \begin{quote}
%     "I think my workflow has become much more efficient because ChatGPT helps a lot so I don't have to search on Google. I can straightaway, go to one platform and get all the answers." -P56
% \end{quote}

\subsubsection{\textbf{Challenges in the Utilization of ChatGPT}}
A majority of the students faced ethical dilemmas while using the tool for their academic needs. Students were very aware of the fact that it is quite easy to use ChatGPT for unethical purposes within academia. Some students mentioned that using ChatGPT felt like an unethical shortcut, and this made them prefer hard work and traditional methods of acquiring information and getting answers over ChatGPT. Some students also reported feeling guilty using ChatGPT. Students reported that they feel ChatGPT is a good assistant to their workflow but it inevitably makes humans more lazy and lethargic. They believed that in the longer run, ChatGPT will have negative effects on the capabilities of the human brain and hence humans face a long-term risk of losing their efficiency. Many students believed that due to the effortless nature of gaining answers and information through ChatGPT, they are learning very ineffectively. They mentioned that using ChatGPT was good for surface-level information, but not for in-depth knowledge. One participant also mentioned that they believe using ChatGPT is letting go of the traditional and rote learning methods of gaining knowledge, which is a negative effect of ChatGPT. Nonetheless, students admitted that despite the long-term risks, the tool is very helpful in the short term.
\begin{quote}
    "Sometimes I feel as though, I am cheating by learning through it. I mean sometimes, you know, it just feels wrong when it gets too easy?" -[P3]
\end{quote}

Trust and reliability-based challenges were one of the most frequently mentioned challenges when it came to students using ChatGPT. The majority of the students mentioned how unreliable ChatGPT is in generating responses, and this reliability has put many students on guard. Some students mentioned how ChatGPT sometimes gives correct answers for specific questions, and sometimes it does not. Students reported varied experiences with ChatGPT responses. It's responses often lacked consistency. Some mentioned that requesting response regeneration resulted in entirely different answers, while others faced persistent repetition of the same response. Paraphrasing also posed issues, with students observing that ChatGPT often changed the original text's meaning and deviated from the original prompt. Moreover, many students mentioned that ChatGPT is incapable of judging whether what it is reporting is right or wrong, and this lack of verification makes it challenging to rely on ChatGPT for learning. 
\begin{quote}
    "I think while paraphrasing, something ChatGPT, completely changed the meaning of those scientific texts. So if I would not have come across and proof read it, I don't think I would have come across the policies and it would have completely proved my Observations wrong." -[P62]
\end{quote}

The extent of help one can seek from ChatGPT is limited due to its accuracy limitations. Additionally, many students mentioned that even if they told ChatGPT where it was wrong, it had an inclination to repeat its mistakes. When it comes to problem-solving, our interviews reported that a majority of students feel that ChatGPT cannot intuitively solve new problems, it generates incorrect code, it is incapable of solving complex numerical-based questions and it lacks in-depth information for specific programs like Mechanical Engineering and Electrical and Electronic Engineering. It also provides outdated information on recent events and is sometimes biased in its views, and portrays information in a very confident manner, even if it is incorrect.
\begin{quote}
    "sometimes the tool tries to give an answers way too confidently about things that it is not really aware about. [...] I found that Political affiliations are a bit questionable. I believe a tool like that should be new impartial and should try to give in fact instead of providing opinion." -[P13]
\end{quote}

Apart from trust and the quality of answers generated, a large percentage of the students also faced general usability-based challenges. A fraction of the students mentioned that making an account and periodically signing in to the service becomes a hindrance to their workflow. Students also mentioned that the chatbot nature of ChatGPT does not feel very humanistic, and feels repetitive in its nature of conversation which makes it difficult to get comfortable with using ChatGPT on a day-to-day basis. Many students mentioned that even if it is good at generating language-based answers, it is not creative enough to meet their creative requirements. A recurring theme throughout the participants was that ChatGPT does not work well on larger, more complex queries. This was followed by many students reporting that in order to cater to their complex queries, they have to break down their prompt into multiple, simpler prompts in order to get the desired answer from the tool. Writing better prompts was a challenge faced by many, and the fact that they needed to manually verify every response drew many students away from the tool. Students were also observed mentioning that the limited database and the restricted usage on the types of questions ChatGPT can answer, and how even slightly controversial topics are avoided by the tool was a pain point. Moreover, many students reported that they wished there was multi-modal access for academic purposes.
\begin{quote}
    "You have to double-check everything. You do it carefully, line by line, and make sure you exactly understand what it's doing. And most of the time, 80\% of the time, you'll have to edit it a lot to get it to work." -[P10]
\end{quote}

\subsubsection{\textbf{Opportunities to Improve - Student Perceptions and Recommendations}}
Across all participants, the majority of participants agreed that ChatGPT has proven to be a helpful tool in their coursework and academic requirements. Many students referred to ChatGPT as an indispensable tool and believed it has had a big impact on education. Students mentioned that there needs to be a balanced usage of the tool to effectively aid their academics, and the tool can be a revolution in education if used appropriately. Some students compared ChatGPT to the internet revolution, and drew parallels between Google Search and ChatGPT, mentioning that ChatGPT is the new Google search and it is here to stay. Some students also expressed their worries when it came to jobs after their undergraduate education, and addressed the opinions that ChatGPT can end up taking jobs. Some students did not believe that ChatGPT would take away their jobs, while others reported having a strong opinion that ChatGPT might wipe away mundane jobs.
\begin{quote}
    "It's a wonderful tool and I think we shouldn't overuse it. We should definitely use it to kind of support our needs or complement them, but we shouldn't rely on it too much." -[P17]
\end{quote}

The majority of students compared ChatGPT with the traditional methods that were in use pre-ChatGPT. The majority of the students mentioned that they feel ChatGPT has the potential to decrease human creativity, while the opposite emotion was also brought up stating that the human mind is much more efficient compared to ChatGPT. Some students felt that the reliance on the human mind has decreased since the introduction of the tool. On the other hand, a few students reported that they felt ChatGPT was not useful enough, and they did not feel any major contributions made by ChatGPT in their academic workflows. 
\begin{quote}
    "I guess it would obviously change the world in a way that we will rely less on humans, and humans \textit{(will rely)} more on ChatGPT." -[P19]
\end{quote}

Many students commented on their usage habits with ChatGPT. Some students reported that it took them some time to get used to the tool and understand how to use it efficiently. Many believed that structuring prompts in the right way was essential to use the tool effectively, and it requires a certain level of expertise to be used correctly. Students mentioned that over time they have learnt when and when not to use ChatGPT. Students reported making clear distinctions of when ChatGPT can be trusted with it's answers. The majority of students believe that ChatGPT is time-saving and convenient to use for academic use cases. Many students mentioned that they feel the help they get from ChatGPT is personalized, and that allows them to get help when other traditional agents of guidance, like instructors, are unavailable. Some students felt that ChatGPT was on a thin line when it came to being good or bad due to it being a trade-off between efficiency and laziness.
\begin{quote}
    "So when I'm talking about learning new things and concepts, I know that I can trust it, but if I ask it to solve somethng for me, I know it can't be that reliable. So I know exactly when I have to use it." -[P30]
\end{quote}

Throughout our interviews, students put out their opinions openly and gave suggestions. Many students reported that they believe ChatGPT should hold itself accountable for the answers it gives, and have admittance to not having an answer instead of giving a vague response. In terms of learning to use the tool, students believed that there should be in-built tutorials to help students realize the potential of the tool, rather than having to explore it on their own. The majority of the students mentioned that proper training and guidance should be provided to some extent in terms of how to use the tool effectively. Students also wanted the tool to be more interactive, with features such as cross-questioning. Regarding academic recommendations, some students mentioned that they would want a ChatGPT model that can deal with research work better, including training the model on research papers, having books and their solutions integrated into the tool, and being able to accept equations, formulas, and diagrams better. Many students also highlighted that they would prefer if the open-access version of ChatGPT (GPT 3.5) supported multiple forms of input, including but not limited to images, mathematical calculations, graphs, and more. Students also wanted the interface to be more customized so that they could tweak the model to their requirements before they started giving prompts.
\begin{quote}
    "I think if ChatGPT was such that when I regenerate my response, then ChatGPT could ask itself, [...] ask me questions. What exactly am I looking for? And then give it's response. That would be great." -[P15]
\end{quote}

\subsection{Instructor Perspective on ChatGPT}
\subsubsection{\textbf{Instructor Awareness about ChatGPT and its uses}}
Instructors learned about ChatGPT through word-of-mouth, news articles, interviews, and online buzz, sparking their interest in trying it out. Some instructors initially viewed it as a potentially \textit{over-hyped} technology and questioned its practical utility. One of the instructors compared ChatGPT to the early days of Google, suggesting that its initial results might not be perfect as was the case with Google. On the other hand, several instructors were impressed and enthusiastic about ChatGPT's capabilities, especially in terms of information retrieval, summarization, and problem-solving. Instructors both praised ChatGPT for answers and highlighted its inaccuracies and limitations, particularly for solving numerical problems. 
% Some expressed concerns about students relying too much on it for quick answers instead of deep learning and comprehension.
\begin{quote}
    "I came across ChatGPT because of the hype it generated on the internet. It was crazy, so like everyone else, I also tried my hand at it." -[T9]
\end{quote}
Coming to student usage, many instructors admitted that they are not fully aware of how students are using ChatGPT in their courses. Some of them did express an intention to initiate discussions with students to gain a better understanding of their experience with ChatGPT. Some instructors also mentioned that the nature of the take-home assignments in their courses is such that they did not think ChatGPT would be able to solve them. On the other hand, some instructors were aware of students trying to use ChatGPT in open-internet exams and assignments. Instructors worried about students blindly relying on ChatGPT without understanding the meaningfulness of its responses, and some even caught students doing so.
\begin{quote}
"Even in the labs, we observed students blindly copying code for tasks related to communication systems, which may be fine. But the problem is that they don't really understand what they're copying. They just want to get the work done. It undermines the learning of the students." -[T16] 
\end{quote}

\subsubsection{\textbf{Instructor Perceptions on Student Learning through ChatGPT}}
Instructors shared that ChatGPT can complement students who are already proficient in a subject, aiding them in learning more about that subject. However, its reliability may not suffice for critical assignments, necessitating students to have a background in the topic to identify inaccuracies. Some instructors also believe that ChatGPT can help in focused and efficient learning by providing concise and specific information. This saves them valuable time that might otherwise be spent searching through numerous sources. Instructors also thought that ChatGPT could simplify complex concepts, making them accessible to students with varying levels of subject proficiency and helping bridge language barriers. Some instructors also mentioned that the ChatGPT platform offers a distraction-free and safe learning environment, devoid of advertisements and provides controlled access to reliable information, particularly beneficial for young learners. The interactive and user-friendly interface of ChatGPT was seen as an enabler of iterative learning and reinforcement. Students can receive assistance, identify mistakes, and improve their understanding through back-and-forth interactions. Additionally, ChatGPT supports experimentation and exploration of \textit{what if} scenarios on a particular topic for deeper comprehension. Instructors also mentioned that students can leverage ChatGPT to generate practice questions on any topic. This can enable students to create their own exercises based on course material, promoting a deeper understanding of concepts. In this context, an instructor highlighted that ChatGPT may sometimes provide incorrect answers to a question. This discrepancy can be used as an opportunity to debate with peers on the correctness of the answer, ultimately, contributing to enhanced learning and a deeper comprehension of course material.
\begin{quote}
    %If students use it efficiently, it can really help them learn things in a very focused manner. 
    "In classroom teaching, we have limited time, so we can't delve into every topic in great detail. At the same time, we don't expect students to go through every topic in depth. [...] 
    %Some topics are crucial, some we want them to skim through, and some we want them to explore with a bit more depth. 
    The advantage of ChatGPT is that it doesn't overload you with a lot of information; it provides very focused information on a specific topic. This can be very satisfying for a student, and they can feel confident that they've understood that topic [...] ChatGPT provides a more streamlined experience \textit{compared to Google Search}." - [T4]
\end{quote}
Instructors viewed ChatGPT as a valuable tool to improve students' writing skills, especially for non-native English speakers like Indian students. It provides suggestions for grammar, structure, and presentation in various types of written communication. Additionally, it offers guidance on research paper writing and can generate templates. Exposure to ChatGPT-generated high-quality written material can also enhance spoken communication, as one instructor noted. Instructors saw ChatGPT as a resource to help students express mathematical ideas more clearly by generating structured outlines for theorem proofs, aiding in effective mathematical proof-writing skills. Many instructors mentioned that ChatGPT can assist students in generating basic code for assignments or projects, serving as a starting point to save time and effort. However, they recommended that ChatGPT should be used for coding assistance only from sophomore year or later so that students can grasp fundamental programming concepts.
\begin{quote}
    "In my courses, we often have to prove theorems and establish various facts. [...] When I provide ChatGPT with an outline of the proof, it can generate a well-structured and elegantly written proof [...] the way ChatGPT formulates the proofs is often superior to what an average second-year undergraduate student could produce. This skill of effectively communicating solutions or proofs is something that everyone should have. Even if you know the answer or solution, being able to convey it clearly and elegantly is essential." - [T13]
\end{quote}
Instructors emphasize that students should understand that ChatGPT is a tool to aid their learning, not a replacement for their intellectual capabilities. It is important for students to use ChatGPT responsibly and not rely on it excessively. For example, ChatGPT could be used for small or trivial parts of assignments or when students need quick assistance, but not as a complete substitute for their own learning efforts. More so, instructors emphasize that students should understand that ChatGPT is a tool to aid their learning, not a replacement for their intellectual capabilities. It is important for students to use ChatGPT responsibly. For example, ChatGPT could be used for small or trivial parts of assignments or when students need quick assistance, but not as a complete substitute for their own learning efforts. More so, 
% \begin{quote}
%     "If students understand that an assignment is meant for them to grasp the concepts, and they use ChatGPT for the right purpose, like for a small, trivial part or to save time on certain aspects, I think that's fine. But if they rely on ChatGPT for the entire assignment, then I believe it's a problem." - [T12]
% \end{quote}  
% Instructors expressed concerns that ChatGPT could tempt students to seek quick answers without investing the time to understand the material truly. This behavior may undermine the learning process as well as academic integrity. An instructor mentioned that students may view copying from ChatGPT as a less immoral alternative to copying from peers. Some instructors were worried about the possible decline in students' intellectual abilities due to over-reliance on ChatGPT, drawing parallels with the technology dependency seen with smartphones and calculators. They feared that ChatGPT may discourage systematic learning, critical thinking, problem-solving, and creativity. Some participants were also concerned that students might become passive consumers of information, diminishing their analytical skills.

\begin{quote}
%    There will always be a threat. I don't know if there was a survey conducted by Harvard a few years ago, where they concluded that with the advent of smartphones, human intelligence quotient is decreasing. I can relate to it personally. When I had a basic phone, I could remember six to seven phone numbers in my mind, ready to dial at any time. But now, I can't even remember my wife's phone number. So, what has happened? 
    "We've become so dependent on technology that we've stopped using that part of our brain. I suspect that if this dependence on technology goes unchecked, it may lead to a situation where human intelligence is compromised. [...]
    % Students tend to seek the path of least resistance. 
    It might reach a point where students say, "Why should I study now? I can do it later with ChatGPT." This attitude could continue and eventually lead to students trying to use it for unfair advantages in tests or exams. So, any technology should be used in moderation and enjoyed responsibly." -[T3]
\end{quote}

Instructors expressed concern that the availability of tools like ChatGPT may reduce students' inclination to seek help from their peers or teachers when they encounter challenges or questions as they may now resort to ChatGPT. This may make the students less social with other students and instructors. Hence, it's important to strike a balance between utilizing technology for assistance and maintaining social interaction and collaborative learning. One instructor highlighted that if everyone relies on the same AI tool such as ChatGPT for generating content or answers, it could lead to homogeneity in thinking and content production. 
\begin{quote}
    "It's gradually reducing in that aspect. In the past, when students didn't know something, they would ask their peers or teachers for the solution or the underlying concept. But these interactions seem to be missing these days, and it's kind of un-socializing the students." -[T6]
\end{quote}
% \begin{quote}
%     "It might pose a threat of bringing about homogeneity in people's thinking because if everyone relies on ChatGPT or a similar tool for answers, they may end up with similar responses. This could lead to a more homogeneous and less diverse range of perspectives and ideas. It's not just a threat but a broader issue." - [T13]
% \end{quote}

\subsubsection{\textbf{Influences on Teaching and Assessment Methodologies}}
Instructors mentioned that they can use ChatGPT to enhance their teaching by finding alternative explanations for challenging concepts and generating additional examples and practice questions. This approach can not only aid in conveying information more effectively but can also provide students with practical applications of the material. ChatGPT's generative ability assists in creating diverse, up-to-date lecture materials and well-structured assessments, increasing student engagement and understanding. Instructors also explained that ChatGPT can help them generate relevant content within seconds, thus, optimizing their teaching efficiency. A computer science instructor suggested that instructors utilize ChatGPT to create interactive simulators for teaching a system or algorithm to provide students with practical, hands-on experiences that help them better understand complex systems. A few instructors mentioned that ChatGPT can also assist in course design by suggesting the most effective sequence of topics and relevant course material to improve learning outcomes and engagement for students. This saves time in researching and designing course content and also ensures that the course remains up-to-date and relevant. Instructors also emphasized the need to embrace technology and innovation. They encouraged other instructors to leverage tools like ChatGPT to enhance the learning experience and recommended becoming familiar with online resource limitations and strategies to overcome them.
\begin{quote}
    "A couple of months ago, I was designing a new course. [...] 
    % While I was preparing for it, 
    My senior colleagues suggested that I should explore other universities' portals to see if they are offering similar courses for their students. This took up a significant amount of my time, and often, I found that the information I came across was not very relevant. ChatGPT, being more optimized, could save a lot of time in designing course content. For example, once I've finalized a course topic and its content, there might be elements that I'm missing, that could add value and are already taught in similar courses at other universities. If ChatGPT can highlight such relevant information, it would enhance the quality and content of our teaching while also saving time." -[T16]
\end{quote}

% A computer science instructor suggested that instructors utilize ChatGPT to create simulators for teaching a system or algorithm. By using ChatGPT to create such interactive simulations and assessments, instructors can enhance their teaching methods by providing students with practical, hands-on experiences that help them better understand complex systems and algorithms. 
% \begin{quote}
%     "Implementing a router using ChatGPT was an interesting experiment. The questions were designed in a way that routing entries were being created, and the system had to determine how to route a given packet type to the correct interface. There were instances where it successfully worked, and there were also cases where it was not successful." -[T14]   
% \end{quote}

However, one instructor was also of the opinion that using ChatGPT to prepare teaching material is cheating and should be avoided completely. They believed that the teaching material should reflect their genuine understanding and interpretation of the subject matter. They believed in the value of building content from the ground up and citing sources transparently, emphasizing the role of their own scholarship in the teaching process.

\begin{quote}
    "Every line I write should reflect something I genuinely believe in, something that has become a part of my understanding. [...]
    % I'm not even copying from textbooks. 
    If I'm not allowed to copy from textbooks, how can I justify copying from ChatGPT? I must think about what the textbooks are conveying, interpret the material, and create my own slides. It should be a product of my scholarship. If I use ChatGPT, it would essentially be cheating. Furthermore, there's a concern about the reliability of sources. [...] As a teacher, I should 
    % start from scratch, begin with a textbook, and 
    build my content from the ground up, citing all my sources in my slides." - [T2]
\end{quote}

Instructors mentioned that they could move towards higher-order learning and problem-solving while asking students to use ChatGPT for more routine and maintainable tasks if ChatGPT gets more pervasive. For instance, in a software engineering course, instructors could focus more on advanced concepts and delegate routine tasks, like writing code modules, to ChatGPT. This would allow students to engage in higher-level thinking and problem-solving. However, the participants also acknowledged the challenges of developing such a higher level of understanding without going through the foundational learning processes. For instance, what to remove from existing content/curriculum to accommodate new higher-level concepts. Additionally, adapting learning outcomes and revising course content assumes a universal reliance on ChatGPT and similar tools, which may not always be true.
\begin{quote}
    % "
    % I lack clarity on this issue. 
    "It appears that we need to move towards a higher order of learning, focusing on complex problems and solutions, while using ChatGPT for lower-level, more routine tasks. However, it's still not clear to me how we can achieve a higher level of complexity without going through the foundational learning process. Can we skip this process entirely? My intuition tells me that we cannot develop a deep understanding without going through the process of grasping the fundamentals. It's like trying to become a proficient coder who assembles modules created by ChatGPT without having a solid understanding of the underlying concepts." - [T11]
    % I'm not entirely sure about the feasibility of that approach.
\end{quote}

Many instructors acknowledged the need to reconsider traditional assessment methods in light of ChatGPT's presence. Instructors recognized that in-class closed-book exams are likely to retain their significance in undergraduate assessment due to the controlled and supervised testing environment. This controlled environment helps maintain the integrity of assessments, making them less susceptible to the influence of AI tools like ChatGPT. However, they also acknowledged the importance of supplementary assessment components, such as take-home assignments and projects, to foster deeper engagement with course material. However, many instructors found open-book or open-internet assessments problematic due to the potential for students to rely on ChatGPT. Some instructors have already removed open-book assessments, while others suggested contextualizing questions to prevent ChatGPT from offering ready-made answers. Assignments that require real observations, personal experiences, and synthesis of information could be designed to promote deeper understanding and discourage reliance on AI. Instructors may also adjust the complexity of problems to make them less amenable to ChatGPT-generated solutions.
\begin{quote}
    "Earlier, I used to assign term papers, but now I've shifted the focus. I've added more weightage to assignments that require observation and practical involvement. Students need to write about their observations and even take real photos as part of their assignments. They can't simply rely on internet sources for everything. 
    % I emphasize these points in my assignments
    [...] This way, I can tell if they've genuinely engaged with the subject matter. I removed the term paper requirement because I know they can easily write it using ChatGPT. Secondly, in design courses, projects, and assignments, most of them are not very binary; they involve observation and presentation, such as creating a video. I try to make them as non-objective as possible, so AI may not be as useful in that context. This doesn't mean I make the assignments overly challenging, but analyzing existing products or designs around them can provide valuable insights." -[T19]
\end{quote}

Instructors also found open-book assessments more challenging for detecting plagiarism due to the difficulty in distinguishing between original student work and ChatGPT-influenced work. This raises concerns about verifying actual student learning during grading. Students may also alter AI-generated answers to appear original. Instructors suggested the need for discussion and brainstorming among peers to address these challenges and explore methods to detect AI tool usage while maintaining fairness and integrity in evaluation.
\begin{quote}
    "It's difficult as an instructor to figure out answers written using ChatGPT. I mean, somebody is writing an answer, and I have a moral and academic responsibility of evaluating their answer and giving them credit if it is original and not giving credit if it is not original. But that is very difficult. I mean, in a non-regulated setting, I don't know if you got it, it's not a very hard task to modify a Chat answer to make it look like it's an original. [...]
    % I think it's an important point as an instructor. 
    So I feel that now when I give homework marks to someone, that 10 out of 10 may not actually mean that the student knows the stuff. I'm not able to verify the learning of the student." -[T10] 
    % There is some brainstorming required or discussion required on how to either change the way we detect cheating or change our style of evaluation.
    
\end{quote}

\subsubsection{\textbf{Instructor Recommendations}}
Instructors suggested that ChatGPT should provide confidence scores with its responses to help users assess the reliability of the information. Some instructors also proposed that ChatGPT should provide references or citations for the specific sources that it has used to generate a particular response, ideally in a journal-style in-line citation format. These features would make ChatGPT more accountable as well as enable users to make informed decisions when consuming ChatGPT-generated content. A few instructors also suggested that users should be encouraged to cite ChatGPT as a source in their research articles, increasing awareness about AI-generated information. Instructors envisioned ChatGPT evolving to provide richer, interactive education with multimedia, including images, graphics, animations, and spoken interactions. This would offer a dynamic, engaging experience beyond text-based interactions. Some instructors recognized the challenges of inputting mathematical queries into ChatGPT and recommended improving the interface for easier input of mathematical formulas and equations. This will make ChatGPT more convenient for users seeking solutions to mathematical problems.
\begin{quote}
"I think there should be some accountability [...] because the problem that I can see is that if ChatGPT is so powerful, and in the future, if people see it as a replacement for Google, then...it can also generate biased information. Who will be accountable for that? When you do Google search, Google itself is not providing any information; it can point to different blogs or something, and you know whether those people are reliable or not. Even some media house can be biased, some can be unbiased, so you can look at different media houses, then you can build your understanding about, like, what is what. But if everything is a black box, and then you're just getting some information, and then that is a bit problematic, so…" - [T12]
\end{quote}

Instructors noted that ChatGPT could serve as a foundation for various new tools and applications. Just as Google provided a window to the world, ChatGPT is seen as a foundational technology upon which new tools can be built. These tools could be domain-specific and tailored to the needs of researchers, students, developers, data scientists, and more. For example, a student-focused version of ChatGPT would not simply provide answers but act as a tutor, giving hints, directions, and clarification. This approach aims to strike a balance between saving students time on redundant tasks and promoting active learning by not providing complete solutions right away. 

\begin{quote}
    "Currently, there is a single ChatGPT and it's based on a universal model. In the long run, probably what we will see, and this might not be good for the company as such, is a lot of chatbots tied to different objectives. There might be a chatbot that is just for helping people with assignments. There might be a chatbot that is designed for instructors to improve their questions." - [T7]
\end{quote}

Instructors expressed concerns about potential bias in the training data of ChatGPT, especially with regard to its applicability in regions that may not have been adequately represented in the data. They suggested that fairness, bias, and transparency should be thoroughly reviewed in the training process. Addressing bias, ensuring representation from diverse sources, and providing transparency and explainability in AI systems are crucial steps to ensure their responsible and ethical use in various contexts, including regions outside the primary training data sources.

\begin{quote}
    "Working in a developing country, [...]
    % also called a third-world country, 
    it is my apprehension that there may not be sufficient fairness in ChatGPT's training. They may not have used a sufficient amount of data from the region where perhaps they could not reach out sufficiently enough. Therefore, it has been trained on data from the US, and that may not directly apply to the real-life cases in our setting. So, my suggestion would be that fairness, in their training, that should be reviewed. What data they have collected, where they have collected from, whether sufficient attention has been paid to fairness and bias, transparency, it is called explainability. " [T9]
    % I think that's kind of a standard thing which everyone is...
\end{quote}

\section{Discussion}\label{sec:discuss}
% llms as agents of academic independence

\subsection{Expanding the Horizons of LLMs in Education: Building on Existing Capabilities}

Our findings reveal numerous existing use cases and advantages of ChatGPT, as reported by both students and instructors. We identify opportunities to enhance these current benefits in order to further enhance the positive impact of Large Language Models (LLMs) in education. These opportunities and recommendations can be leveraged by designers, engineers, and instructors engaging in the space of LLMs for education. Below, we provide a categorization of these benefits and their potential implications.

\subsubsection{\textbf{LLMs as facilitators of learning and skill building}}

Our findings highlight that both students and instructors acknowledge the significant role ChatGPT has played in supporting learning and the conceptualization process for students. Practices of ChatGPT being used for learning and comprehension included the generation of examples and alternate explanations through ChatGPT for easy understanding, simplification of complicated concepts taught in courses, experimentation, and exploration of new topics, generation of practice questions, and getting feedback on adopted approaches. ChatGPT also came up as an \textit{agent of academic independence} as using ChatGPT can minimize dependence on instructors and peers and their availability as shown in our findings. ChatGPT can provide students with a quick and always available source of information-seeking and doubt-solving. Instances of ChatGPT being used in parallel to in-class learning and beyond-classroom learning were commonly talked about. Instructors also talked about the distraction-free and interactive nature of ChatGPT which enables a safe learning, engaging yet focused environment for students. ChatGPT being used to enhance language skills like proficiency in English is also a significant use case in contexts like India where English is not a native language but most course materials are in English. Exposure to high-quality written material generated by ChatGPT can also contribute to improved spoken communication in the English language.
\\
These use cases open up opportunities to leverage them and build technology-enabled learning and teaching tools over them. We observe that ChatGPT is commonly used for course-specific learning and doubt-solving. AI-powered learning tools and plugins can be built using LLM APIs that can be tailored to specific student needs. Prior literature has extensively explored the use of chatbots to educate students \cite{yeh_how_2022, lee_i_2020, mukherjee_impactbot_2023, gibellini_ai_2023}. The improved and unprecedented capabilities of recent LLMs allow them to adapt better to specific domains and requirements. Tutoring applications built on these LLMs can be specified to serve as personal tutors to students for different subjects. These tools can enable iterative and reinforcement learning of skills relevant to their coursework as well as essential interpersonal skills. Prior literature has explored various ways of incorporating interactive quizzes in learning environments as they serve as a quick way of reinforcing learned concepts \cite{gibellini_ai_2023, lu_readingquizmaker_2023}. Such self-assessment quizzes based on the topics searched and queried can be generated by LLM-based learning tools or plugins and prompted to the learner amidst their usage. Users can also be prompted to self-report if they understood the queried theme correctly. Based on user's inputs, they can be provided with more personalized examples and variations of responses to match their learning requirements and pace. Recommendations of topics or questions similar to queried themes can also be presented to the learner to enhance their knowledge and spark curiosity. More so, multimodal outputs including relevant images or videos can be incorporated in solutions provided by the LLMs to make concepts easier to understand. Better training on regional languages can help in adoption across students coming from educational mediums apart from English.
\\
Additionally, LLM-based tools can be designed to aid instructors in providing more detailed and personalized feedback to students. Such tools can consist of features that let instructors provide information on evaluation criteria and rubrics of feedback to the LLM. This feedback can consist of detailed explanations of student's mistakes and ways to improve. LLM-based tools can also help students practice/rehearse for the in-person viva, presentations, interviews, etc. with personalized feedback. Students can also practice language skills by simulating conversations with LLM-powered teaching bots. Students can feed personal goals in such applications and get feedback specific to their goals. The conversational nature of LLMs allows students to correct their responses, which can help endorse interactive learning. The interaction can also be a Voice User Interface by including speech inputs and outputs to enhance the conversational nature of the interaction \cite{kim_i_2020, reddy_making_2021}. Massive open online courses (MOOCs) are also widely used by students for online learning and accessing courses that might not be accessible to them offline. These MOOCs can use LLM-powered teaching assistants that can solve subject-specific doubts. They can be trained to understand video content so the student can ask doubts about the video content and seek one-on-one help quickly and increase retention \cite{zheng_retention_2015}.

\subsubsection{\textbf{LLMs as an Aid to Academic Tasks}}
Our findings highlight the numerous instances of students using ChatGPT to complete their assignments. Some instances involved using ChatGPT as an ideation tool and a knowledge bank to refer to for assignments. The speedy assistance offered by ChatGPT was a major factor in relying on it for academic assistance. Students mentioned that the near-perfect summarisation and content generation of ChatGPT really helped save time and effort. The instructors mentioned that ChatGPT can make their workflows more efficient as it can also be used for creating diverse, up-to-date lecture materials, well-structured assessments, interesting conceptual simulations, and examples to provide a more interactive and efficient learning experience for students. Moreover, ChatGPT has also been used as a task management tool by students for generating schedules and timetables and creating course roadmaps.
\\
Prior literature has studied task management processes and how technology can ease this process \cite{kamsin_personal_2012, toxtli_understanding_2018}. LLMs can help change the way task management is done owing to their adaptive conversational abilities. Task management applications for instructors and students can be created where they can be asked to feed their tasks, goals, and personal preferences. A collaborative process between the user and the LLM can help efficiently create workflow while LLMs can provide additional tips and strategies to manage tasks. More so, LLM-powered curriculum planners can be developed. Previous work has studied and recommended curriculum practices for teachers to devise effective curricula \cite{lin_how_2021, liu_visual_2023}. LLMs fine-tuned on such guidelines and education literature can be used as assistants for instructors while designing curricula. Instructors can query such tools for specific recommendations and suggestions to make course activities more interesting. Recent work has also explored the possibility of LLMs to mimic user personas \cite{hamalainen_evaluating_2023}. Such functionalities can be leveraged to test course activities on LLM simulations to predict how students would engage with the activity and if it can meet the intended goals of the course. Content generation capabilities of LLMs can be better leveraged by creating special content authoring tools over LLMs. Such tools can provide specific features that can help users change the tone and nature of the content and ideas generated through a Visual Interface instead of textual prompts.

\subsection{Navigating Challenges through a Learning-Focused Design}
Through our findings, we observe different types of challenges, concerns, and perceptions that LLMs pose for students and instructors. Building upon those, we focus on identifying such effects and exploring how future developments can utilize our contributions to develop improved solutions for academia.

\subsubsection{\textbf{The Dilemma of Academic Usage of LLMs}}
Our findings suggest that both students and instructors face an ethical dilemma when it comes to utilizing the various use cases that LLMs offer, compared to the more traditional forms of academic learning practices. Introducing open-access LLMs like ChatGPT raises concerns regarding the use of such tools to facilitate unjust and unethical practices such as students committing plagiarism in their assignments, reports, exams, and other forms of academic evaluations. Moreover, many students expressed similar sentiments, with instructors worrying that using LLMs to copy information might be perceived as a \textit{less immoral} practice than copying from peers. Students also mentioned their preference for hard work, and how they feel that they learn more when they make use of traditional methods of acquiring information. Furthermore, the inherent accuracy and user-trust issues with LLMs like ChatGPT act as a roadblock to the smooth integration of such models, and our findings suggest that these were the most common challenges. Lastly, instructors suspected that using ChatGPT-like tools can reduce the development of cognitive skills like critical and systemic thinking, problem-solving, and creativity among students as they might \textit{over-rely} on it for direct and easy answer-seeking. instructors also expressed concerns about ChatGPT resulting in \textit{homogeneity} among student thinking and approaches. The usage of ChatGPT-like tools was also feared to reduce social interactions of students to discuss and collaborate among peers.
\\
We acknowledge the potential harm LLMs could impose on student learning as they ease up the process of knowledge creation and could be used unethically. We emphasize the need to prioritize \textit{Responsible Learning} in both the design of LLMs for educational use cases as well as the curricula developed for undergraduate students. Students need to be taught about the responsible use of AI. A transformation like that of LLM-enabled learning will also require reforms in educational structures \cite{dignum_role_2021}. AI Ethics education should be made a core part of the undergraduate curriculum to educate students about the capabilities of AI and foster a responsible attitude toward the use of AI. Studies have explored the use of gamification to help individuals identify ethical concerns in the design of AI technologies \cite{ballard_judgment_2019, elsayed-ali_responsible_2023}. Such interactive elements can help students engage in hands-on experience of identifying the ethical use of AI. Designers and developers of LLM-based educational tools should also be educated about practices that can be adopted to create learning-friendly interfaces. Various studies on Responsible AI have emphasized the importance of participatory design for building responsible AI applications \cite{rakova_where_2021, schiff_principles_2020, polak_teachers_2022}. Focused group studies can be carried out with instructors and students to recognize how they would visualize a safe LLM-powered educational tool. These studies can help create mental models of how users will make use of LLM-powered educational tools and inform the design of such tools. More so, it remains essential to re-imagine how students are evaluated and tested in undergraduate education. With the introduction of LLMs, the need for holistic evaluations that test and foster students' individual creativity becomes important. Panel discussions between instructors can help facilitate the design of curricula in ways that incorporate LLMs as assistant tools for students' learning. Students can be evaluated on the creative process of solving assignments and asked to provide multi-modal deliverables like videos, digital photo essays \cite{noauthor_photo-essay_2023}, and digital process books. Digital tools and applications for students can be built to support the creation of such deliverables. More so, LLM tools for education can be designed in ways that help students build concepts constructively. Explanations and generations can be broken down through in-context learning \cite{kossen_-context_2023} or prompt engineering that directs the LLM to break answers into steps while prompting the user to interpret and apply the explained concepts in order to proceed to the next steps. Moreover, explainability features in LLM interfaces can help students understand and follow the reason behind the generated outputs which can enable greater trust in the responses \cite{zhang_towards_2022, wang_designing_2019}. Confidence scores or markers can also help students know the authenticity of generated answers.

\subsection{Interaction with LLMs}
Our findings highlight some challenges our participants faced while interacting with ChatGPT. These challenges are common across most LLMs as the interaction modalities for them remain similar. Concerns regarding making an account and having to log in regularly came up as a hindrance in the information-seeking process. Concerns around the inability to rightly prompt to get desirable results were commonly reported. More so, concerns about the robot-like answer generations also came up. Findings suggest that many found that conversing with ChatGPT lacked humanness at times due to factors like the repetitive nature of its generations.

The LLM technology is developing at a fast pace and we acknowledge that various technical challenges that were mentioned by our participants might eventually be addressed. Nonetheless, there is a need to provide students and instructors with resources to learn strategies for prompting LLMs as that is the primary way to collaborate with LLMs. Prompt Books \footnote{https://dallery.gallery/the-dalle-2-prompt-book/} have come up as a way of training users about prompt engineering for generative AI models. These prompt books can be made multimodal by incorporating gamification or conversational elements. Videos and simulations can also serve as training materials for methods of prompting. Inbuilt plugins and features can also suggest alternate prompts to users or help with tips to prompt rightly. MOOCs (Massive Open Online Courses) or social media channels can be leveraged to share more such techniques and strategies. The information-seeking process can be made simpler and direct by the development of plugin extensions that can help summon LLMs-based chatbots across platforms in the form of popups to seek instant information and query resolution. Users should be given accessible features to provide feedback on the outputs generated which can be utilized to personalize the nature of generations to the preference of the users.

\section{Conclusion}\label{sec:conclusion}
This paper presents an analysis of students' and instructors' perspectives on ChatGPT usage within undergraduate engineering programs in Indian universities. Our research incorporates data from 1306 student surveys, 112 student interviews, and 27 instructor interviews across three Indian universities. Our study highlights the potential of ChatGPT like LLMs to reform the academic practices of both teaching and learning. However, there are certain regulatory measures need to be placed to protect students from harming their learning and development. Our findings and recommendations can be used to inform the design of future educational technologies built to assist students and instructors. We also call for dialogues and conversations about reforming traditional teaching methods to adapt educational environments with technological advancements, to enable learning to be enhanced and augmented by these innovations.
% We find that students are employing ChatGPT for multiple academic purposes, including content generation, doubt resolution, learning new subjects, addressing programming issues, and interview preparation. However, students do acknowledge that excessive reliance on ChatGPT may diminish their self-reliance and critical thinking skills. Among instructors, there is a split opinion: some view ChatGPT positively as a tool to enhance learning and teaching methods when used responsibly, while others strongly oppose its use in academic settings, citing potential harm to both students and instructors. Building upon our findings, we discuss how LLMs can be used to build learning and teaching tools that prioritize responsible learning.
% \section{Acknowledgments}
\bibliographystyle{ACM-Reference-Format}
\bibliography{chatgpt-ref}

\appendix

\end{document}